\documentclass[aps,pr,twocolumn,showpacs,superscriptaddress,groupedaddress,nofootinbib]{revtex4}  

\usepackage{longtable}
\usepackage{graphicx}  
\usepackage{ulem}   
\usepackage{comment} 
\usepackage{datetime}
\usepackage{dcolumn}

\usepackage{amsxtra}
\usepackage{euscript} 
\usepackage{bm}        
\usepackage{tabularx}
\usepackage{float}
\restylefloat{table}

\usepackage[cmtip,arrow]{xy}
\usepackage{pb-diagram,pb-xy}

\usepackage{amsmath}
\usepackage{amssymb}
\usepackage{amsthm}
\usepackage[pdftex]{color}
\usepackage[pdftex,colorlinks,citecolor=blue,linkcolor=blue,urlcolor=blue]{hyperref} 
\usepackage{graphicx}
\usepackage{dcolumn} 
\usepackage{bm} 

\usepackage{longtable}

\usepackage{ulem}   
\usepackage{comment} 
\usepackage{datetime}

\usepackage{tabularx}
\usepackage{float}
\restylefloat{table}

\newcommand{\beq}{\begin{equation}}
\newcommand{\eeq}{\end{equation}}
\newcommand{\bea}{\begin{eqnarray}}
\newcommand{\eea}{\end{eqnarray}}
\newcommand{\barr}{\begin{array}}
\newcommand{\earr}{\end{array}}

\long\def\begincomment#1\endcomment{}

\newcommand{\Tr}{\mathrm{Tr}}

\usepackage{graphicx}

\usepackage{pgf,tikz}
\usetikzlibrary{arrows}

\usepackage{bm}
\usepackage{bbold}

\usepackage{mathtools}

\DeclarePairedDelimiterX\braket[2]{\langle}{\rangle}{#1 \delimsize\vert #2}

\begin{document}



\title{  Generalized scale behavior and renormalization group for data analysis}

\author{Vincent Lahoche} \email{vincent.lahoche@cea.fr}   
\affiliation{Commissariat à l'\'Energie Atomique (CEA, LIST),
 8 Avenue de la Vauve, 91120 Palaiseau, France}

\author{Dine Ousmane Samary}
\email{dine.ousmanesamary@cipma.uac.bj}
\affiliation{Commissariat à l'\'Energie Atomique (CEA, LIST),
 8 Avenue de la Vauve, 91120 Palaiseau, France}
\affiliation{International Chair in Mathematical Physics and Applications (ICMPA-UNESCO Chair), University of Abomey-Calavi,
072B.P.50, Cotonou, Republic of Benin}

\author{ Mohamed Tamaazousti } \email{mohamed.tamaazousti@cea.fr}   
\affiliation{Commissariat à l'\'Energie Atomique (CEA, LIST),
 8 Avenue de la Vauve, 91120 Palaiseau, France}

\begin{abstract}
\begin{center}
\textbf{Abstract}
\end{center}
Some recent results showed that renormalization group can be considered as a promising framework to address open issues in data analysis. In this work, we focus on one of these aspects, closely related to principal component analysis for the case of large dimensional data sets with covariance having a nearly continuous spectrum. In this case, the distinction between ‘‘noise-like" and ‘‘non-noise" modes becomes arbitrary and an open challenge for standard methods. Observing that both renormalization group and principal component analysis search for simplification for systems involving many degrees of freedom, we aim to use the renormalization group argument to clarify the turning point between noise and information modes. The analogy between coarse-graining renormalization and principal component analysis has been investigated in [Journal of Statistical Physics, {\bf 167}, Issue 3–4, pp 462–475, (2017)], from a perturbative framework, and the implementation with real sets of data by the same authors showed that the procedure may reflect more than a simple formal analogy. In particular, the separation of sampling noise modes may be controlled by a non-Gaussian fixed point, reminiscent of the behaviour of critical systems. In our analysis, we go beyond the perturbative framework using nonperturbative techniques to investigate non-Gaussian fixed points and propose a deeper formalism allowing going beyond power-law assumptions for explicit computations.\\

\noindent
\textbf{Key words :} nonperturbative renormalization, field theory, big data, principal component analysis.
\end{abstract}

\pacs{05.10.Cc, 05.40.-a, 29.85.Fj}

\maketitle
\section{Introduction} \label{sec1}
The physical description of systems involving a very large number of interacting degree of freedom remains a difficult task since the discovery of microscopic structures at the beginning of the 20th century. Since the last decade, the big data revolution has provided a new example of such a large dimensional system, the number of degrees of freedom involved in some data sets can easily be of the same order of magnitude as large dimensional physical systems. Note that the difficulty does not come especially from the size of the system, but from the complex relations between degrees of freedoms. Indeed, there is no difficulty to describe a very large number of independent and identical systems, the complexity reduces to the description of a single one of these subsystems \cite{Avdoshkin:2019trj}-\cite{Hattori:1987jm} and references therein. However, such a reduction breakdown for a system having some complex relations between subsystems; it is, for instance, impossible to reduce the hard complex relations between human cells as the behavior of an isolated cell. More than an impossibility, such a reduction should be a profound misconception. The same thing occurs for data sets, where strong correlations may happen between a vast set of high dimensional data. The principal component analysis (PCA) is based on a systematic dimensional reduction over a finite (and not so big) dimensional space corresponding to the relevant vectors, in the full configuration space, for a given covariance matrix describing correlations between degrees of freedom
\cite{pca0}-\cite{Woloshyn:2019oww}. To be more concrete, a suitably mean-shifted set of data generally takes the form of a big $p\times n$ matrix $X=\{X_{ia}\}$, for $i=1,\cdots, p$ and $a=1,\cdots, n$. Generally, $p$ and $n$ are both assumed to be large, in such a way that the ratio $p/n$ remains fixed. The integer $p$ corresponds to the size of the data, whereas $n$ denotes the size of the set. Then, the covariance matrix $\mathcal{C}$ \textit{between data} is a  $n\times n$ matrix defined as the average of $X^TX$. Orthodox PCA works well when a relatively small number of discrete eigenvalues distinguish from the rest of them, allowing to project linearly onto a reduced dimensional subspace. However, it is not generally the case, as pointed out in \cite{Woloshyn:2019oww}-\cite{Foreman:2017mbc}, especially for a system with a very large number of degrees of freedom, for which the spectra tend to be continuously distributed. In such a case, the separation between relevant and irrelevant eigenvalues becomes arbitrary. This is this arbitrariness that motivates the renormalization group (RG) analysis. \\

The RG is a systematic procedure allowing to describe how a physical system changes at different scales. Originally appeared in field theory as a consequence of the renormalization procedure introduced to solve the problem of ‘‘ultraviolet" divergences, the RG has been essentially developed in the context of critical phenomena. Rapidly, the RG has become a general framework to describe physical systems having a large number of interacting degrees of freedom, and have been used as a powerful investigation tool in a large variety of physical contexts. In particular, in the context of critical phenomena, RG has been proved to be very appropriate to discuss the question of universality. For some critical system, and for a sufficiently large scale, the flow becomes dragged toward a finite-dimensional subspace corresponding to the marginal and relevant operator, providing an efficient projection into a finite-dimensional subspace. Such phenomena, called \textit{large river effect} involves generally a non-Gaussian fixed-point, toward which the mainstream goes from the Gaussian fixed point \cite{KADANOFF:1967zz}-\cite{Pawlowski:2015mlf}. Such a picture provides a qualitative illustration why how RG work so well to describe macroscopic properties of large systems; the dimensional reduction provides an efficient description involving a few sets of parameters, so far from the original complexity of the system. In the physical words, macroscopic physics becomes insensitive to the detailed microscopic structure of the interactions. This is the link that the authors in \cite{pca0} have explored to meet PCA and RG, in the context where PCA introduce an arbitrariness for the choice of the cut-off between relevant and irrelevant parts of the covariance spectrum. More precisely, the authors argued that RG can be used to distinguish between a large number of degrees of freedom those which are sensitive to a change of scale and those that are insensitive; and they propose a field theoretical model to implement this idea.
\medskip

\noindent
The proposed framework was essentially focused on perturbation theory through a partial integration strategy of modes with small variance, essentially inspired from Wilson-Kadanoff point of view on RG. In this way, the authors essentially focused on dimensional effects, their fixed being very reminiscent of the well known Wilson-Fisher fixed point for critical systems. As a consequence, they pointed out that the presence and the relevance of the fixed point depend on the shape of the eigenvalue distribution $\rho(\lambda)$ for the covariance matrix.  The reason for this limitation to the perturbative region is that the authors were essentially interested in distributions not so far from the Gaussian case. However, the analysis that we provide here show that such an approximation cannot be suitable for realistic data sets, and in particular for small deformations around Marchenko-Pastur law, for which the scaling dimensions for couplings are positives and larges, justifying the nonperturbative treatment that we propose in this paper. The manuscript is therefore   written  in a pedagogical style and  serve as an introduction for a series of incoming works for an interdisciplinary community. 
\medskip

The outline is the following: In section \eqref{sec2} we give the useful ingredients and definitions allowing us to compute the FRG equation. Particularly we provide the method used to analyse the scaling behaviour previously given in \cite{pca0}. In section \eqref{sec3} we study the nonperturbative behaviour of the model through the Wetterich equation. We also give the rigorous way to generate the canonical dimension allows deriving the scale variation of the coupling and wave function.
In section \eqref{sec4} the flow equation is solved in the local potential approximation. The numerical investigation and the flow diagrams are also given. Section \eqref{sec5} is devoted to the numerical data analysis which helps to understand the relevance of our method compared with the reference \cite{pca0}. In the same section, we provide our conclusion and remarks.

\section{Preliminaries} \label{sec2}
In this section, we provide some technical aspects useful for the reader in the final steps of the paper. We remain voluntary close to the vocabulary used in the referenced paper \cite{pca0}, to make contact more easily with the result of the authors.

\subsection{Renormalization group for data analysis}
Let us consider a physical system involving a large set $\Phi$ of $N$ random variables $\Phi=\{\phi_1,\cdots, \phi_N\}$, described by a certain probability distribution :
\begin{equation}
p[{\Phi}] = \frac{1}{Z}\,e^{-\mathcal{H}[{\Phi}]}
\end{equation}
where, in accordance to physical nomenclature we call \textit{Hamiltonian} the functional $\mathcal{H}[{\Phi}]$ and \textit{partition function} the normalization $Z$:
\begin{equation}
Z=\int d\Phi \, e^{-\mathcal{H}[{\Phi}]} \,,
\end{equation}
where $d\Phi\equiv d\phi_1d\phi_2\cdots d\phi_N$. It is then useful to define the \textit{generating functional} as follow:
\begin{equation}
Z[J]=\int d\Phi \, e^{-\mathcal{H}[{\Phi}]+\sum_i j_i\phi_i}\,,
\end{equation}
which is such a way that the \textit{connected correlation functions} $\langle \phi_i\phi_j\cdots\phi_k\rangle$ are obtained from $\mathcal{W}[J]:= \ln Z[J]$ by simple derivative with respect to the source $J=\{j_i\}$:
\begin{equation}
\langle \phi_i\phi_j\cdots\phi_k\rangle= \frac{\partial\partial{\cdots}\partial}{\partial j_1\partial j_2\cdots \partial j_k}\, \mathcal{W}[J]\bigg\vert_{J=0}\,.
\end{equation}
Note that $Z[J]$ is usually interpreted as a sourced version of the partition function $Z$, $J$ being interpreted for instance as an external field, bounced the system toward a preferred configuration. As recalled in the introduction, the RG is a formalism allowing to deal with the change of a physical system when the scale changes. In other words, RG supposes a definition of what is microscopic and what is macroscopic. Such a distinction is generally easy for a physical system, like a block of matter; the microscopic scale is identified as the atomic scale and the macroscopic scale to the scale of the magnet itself, involving a large number of microscopic degrees of freedom. For some categories of data, we can eventually define the corresponding notions. For the sets of images, for instance, the microscopic degrees of freedom can be identified with pixels, whereas the macroscopic structures become the ordinary objects ``planes, cats, cars\rq{}\rq{} and so on. However, the situation is not always adequate; and for more abstract sets of data, it may be difficult to identify what is microscopic. In a more abstract level, it is tempting to associate the microscopic level with noisy degrees of freedom; i.e. with the region of small eigenvalues of the covariance matrix spectrum. The RG framework that we discuss in this paper moreover provides a canonical notion of what is microscopic and what is macroscopic.

\begin{figure}
\includegraphics[scale=0.8]{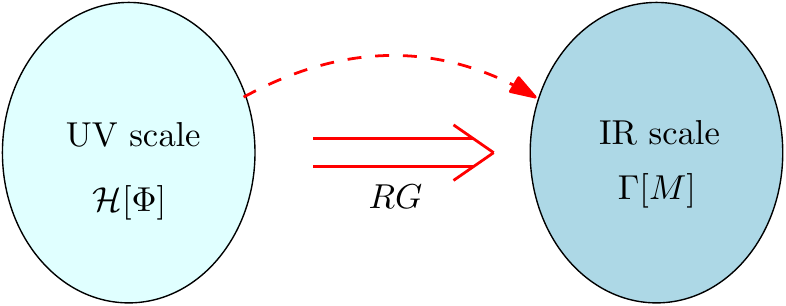}
\caption{The canonical definition of UV and IR scales from the integration point of view. In the UV scale, no degree of freedom is integrated, whereas they are all integrated out in the IR scale. RG provides then a path through scales, from UV to IR. }
\end{figure}

\noindent
Indeed from the RG point of view, at the microscopic level, that we usually name as ultraviolet (UV) scale in physics, no fluctuations are took into account, and the distribution $p[\Phi]$ is essentially dominated by \textit{classical configurations} $\Phi_{uv}\equiv \{\phi_{i,uv}\}$, corresponding to the minima of the hamiltonian functional:
\begin{equation}
\frac{\partial\mathcal{H}}{\partial \phi_{i,uv}}=0 \qquad \forall i\,.
\end{equation}
In contrast, in the macroscopic level, referred to as an infrared (IR) scale, all the degrees of freedom are integrated out. The effective distribution is then described in terms of the effective hamiltonian $\Gamma[M]$, defined as the Legendre transform of the generating functional $\mathcal{W}[J]$:
\begin{equation}
\Gamma[M]=\sum_i j_i m_i-\mathcal{W}[J]\,,
\end{equation}
assuming convexity of $\mathcal{W}[J]$. The effective classical field $M=\{m_i\}$ being defined as:
\begin{equation}
m_i := \frac{\partial \mathcal{W}[J]}{\partial j_i}\,.
\end{equation}
The RG transformations provide a path between these two extreme scales, build as a chain of partial integrations of microscopic fluctuations. More precisely, the Wilson-Kadanoff perspective supposes the existence of slicing in the configuration space of elementary degrees of freedom; such that for each step along the chain, the integration over elementary degrees of freedom into a given slice provide a new effective distribution and a new definition of the UV scale. To be more technical, they assume the existence of finite partitions $\nu_I [\Phi]$ defining a slicing such that $\nu_I \cap \nu_J=\emptyset$\footnote{To be more precise, in concrete example, the partitions $\nu_I$ turn to be distributions over a finite range of wavelength, and may have non-vanishing but small covering between them.}. for $I\neq J$ and that $\forall i$ there exist one and only one $I$ such that $\phi_i \in \nu_I[\Phi]$; and $\Phi=\sum_I \nu_I [\Phi]$. From these partitions, effective distributions $p_1, p_2, \cdots$ may be obtained from integration over each slice:
\begin{equation*}
p_{I}[\Phi_{I}]=\frac{e^{-\mathcal{H}_I[\Phi_{I}]}}{Z_I} \,,\quad p_{I+1}[\Phi_{I+1}] \propto \int d\nu_I [\Phi]\, e^{-\mathcal{H}_I[\Phi_{I}]} \, ,
\end{equation*}
where $d\nu_I [\Phi]:=\prod_{i\vert \phi_i\in \nu_I [\Phi]} d\phi_i$ and $\Phi_{I}:=\sum_{J\geq I} \nu_J[\Phi]$. In terms of hamiltonians, the RG procedure builds a chain of effective hamiltonian $\mathcal{H}_I$:
\begin{equation}
\mathcal{H} \Rightarrow \mathcal{H}_1\Rightarrow\mathcal{H}_2\Rightarrow\cdots\,.
\end{equation}
This is along this chain that the relevance of some operator becomes of crude importance; for a sufficient number of steps, only relevant operators survive, and the microscopic irrelevant details are erased. However, even if we dispose of canonical notions for UV and IR, it does not necessarily exist a preferred path to join UV and IR, and the results may depend on the path that we consider, i.e. on the arbitrary ordering of the elementary fluctuations in the partial integration chain. As explain before, the UV scales have to correspond with degrees of freedom associated with small eigenvalues of the covariance matrix spectrum; and this intuition is recovered in the field theoretical embedding proposed in \cite{pca0}. The idea is to interpret the variance matrix $\mathcal{K}:=\mathcal{C}^{-1}$ as the kinetic kernel for the $N$ variables $\phi_i$ for $p[\Phi]$. With this respect, the simpler distribution we can think about is the Gaussian one:
\begin{equation}
\mathcal{H}[\Phi]= \frac{1}{2} \sum_{i,j}\, \phi_i \mathcal{K}_{ij} \phi_j\,,
\end{equation}
the matrix $\mathcal{K}$ being symmetric by construction. The covariance matrix is interpreted as the second-order cumulant of the Gaussian distribution\footnote{The ‘‘vacuum" $2$-point function in the theoretical physics language.}.\\
Finally, the RG can be constructed as for any field theory, starting by integrating out the degrees of freedom with higher variance. In other words, the spectrum of the matrix $\mathcal{K}$ gives a canonical size for the fluctuations and provides a canonical path to join UV and IR scales toward less noisy degrees of freedom. Obviously, for a purely Gaussian distribution, such a procedure is without interest, the Gaussian distribution is a fixed point of the RG. More interestingly is to evaluate the behavior of perturbations to pure Gaussian distributions, and this is the point we discuss in this paper.
To be more concrete, and following \cite{pca0}, we denote as $u^{(\mu)}_i$ the set of normalized eigenvectors of $\mathcal{K}$, with eigenvalues $\lambda_\mu$,
\begin{equation}
\sum_j \mathcal{K}_{ij} u^{(\mu)}_j = \lambda_\mu u^{(\mu)}_i\,,\quad \sum_{i} u^{(\mu)}_iu^{(\mu^\prime)}_i=\delta_{\mu\mu^\prime}\,,
\end{equation}
we straightforwardly deduce that the Gaussian part $\mathcal{H}_G[\Phi]$ reduces to $\mathcal{H}_G[\Phi]=\frac{1}{2}\sum_\mu \lambda_\mu \,\psi_\mu \psi_\mu$, with $\psi_\mu:= \sum_i \phi_iu^{(\mu)}_i $; and each elementary fluctuations have size:
\begin{equation}
\langle\psi_\mu \psi_\mu \rangle \sim \frac{1}{\lambda_\mu}\,.
\end{equation}
Then, in the most general case, for a non purely Gaussian distribution, the fluctuations can be integrated out following their proper size given by the non trivial spectrum of the matrix $\mathcal{K}$. We denote as $\lambda_0$ the smallest eigenvalue $\lambda_\mu \geq \lambda_0$\footnote{Which is the larger eigenvalue of $\mathcal{C}$.}. Anticipating on the next section, and in accordance with the field theoretical language, we define the square \textit{momenta} $p_\mu^2$ as $p_\mu^2 :=\lambda_\mu-\lambda_0$. The notation reflect the fact that, the matrix $\mathcal{K}$ being positive definite by construction; all the eigenvalues have to be positives $\lambda_\mu \geq 0 \, \forall \mu$ and $p_\mu^2 \geq 0$. The isolated eigenvalue $\lambda_0$ play the role of a mass; and the Gaussian part of the hamiltonian takes the form:
\begin{equation}
\mathcal{H}_G=\frac{1}{2} \,\sum_\mu\, \psi_\mu(p_\mu^2+\lambda_0)\psi_\mu\,.\label{HG1}
\end{equation}
For the reader familiar with quantum and statistical field theory, the previous relation is very reminiscent of the standard kinetic action for scalar field theory, with (square) mass equal to $\lambda_0$. \\

\subsection{The model} \label{themodel}

A coarse-graining in information theory reduces information from each step. This has for consequence that (in absence of critical lines), the divergence (or mutual entropy) between two distributions has to decrease along with the flow. Thus, RG formalism allows to make contact with the intuition that arbitrariness of the cut-off between noise and information should be connected with a partial integrating process over some microscopic degrees of freedom; in such a way that only a few numbers of parameter survive after some steps. The fact that RG concern non-Gaussian distributions are clear by constructions. Gaussian distributions are stable for each step of the RG map. In technical words, the Gaussian fixed point corresponds to a fixed point of the theory, and the distribution remains self-similar at each intermediate scales. Then, the first question is: Is the Gaussian distribution stable for a noisy signal? In other words, assuming a little deviation from the Gaussian fixed point, this perturbation must decrease or increase along with the RG flow? As we will see below, the answer is no for the Marchenko-Pastur distribution. However, it seems clear from the discussion of the previous section that the answer to this question has to depend on the shape of the spectral distribution $\rho(\lambda)$. For this reason, we can hopefully do detect the presence of information in a signal around a noisy distribution from the universal properties of distributions in the deep IR. These properties can be for instance the number of independent correlations functions requires to parametric them, up to the experimental threshold. A physical example of the role played by the shape of the distribution in the asymptotic behavior of the distributions provided by ferromagnetic metals. Here, the shape of the momenta distribution is $\sim (p^2)^{D/2-1}$. For space dimension $D>4$, the Gaussian fixed point is stable, and Gaussian distributions remain valid to describe the asymptotic behavior of the ferromagnet below the Curie temperature. In contrast, for $D<4$, the Gaussian fixed point becomes unstable, and an interacting theory is required to describe the ferromagnetic transition. Interestingly, such a change of behavior has been stressed in \cite{pca0}, investigating the behavior of the normalized $4$-point function $\langle \phi_i^4\rangle/\langle \phi_i^2\rangle^2$ by gradually integrating out degrees of freedom. observed that the behavior of the normalized $4$-point function is drastically modified when some percents of the higher eigenvalues of their spectra are suppressed, and this is what we will formalize in this paper. \medskip

\medskip

In \cite{pca0}, the authors especially focused on the following truncation into the theory space:
\begin{equation}
\mathcal{H}[\Phi]:= \frac{1}{2} \sum_{i,j}\, \phi_i \mathcal{K}_{ij} \phi_j+\frac{g_1}{4!}\, \sum_i \phi_i^4 \,, \label{truncation1}
\end{equation}
and showed that, following the choice of the distribution $\rho(\lambda)$ for the eigenvalues of $\mathcal{K}$; the interaction part may increase with the number of RG steps. The origin of this behaviour can be traced from the scaling itself. The intermediate scales between deep UV and deep IR are fixed by the size of the eigenvalues $\lambda_\mu$. In other words, they provide a scaling, and $g_1$ may acquire a specific dimension concerning this scaling. This dimension dictates how the coupling constant $g_1$ is sensitive to the change of scale. In the RG process, this change of scale, in practice, have to correspond to a change of the cut-off $\Lambda$ corresponding to the upper bound of the spectrum $\lambda_\mu \leq \Lambda$. Then, the natural unit along the increasing scales is the one of $\Lambda$, and the canonical dimension $d_1(\Lambda)$ of $g_1$ is defined as its proper dependence under a dilatation of $\Lambda$; so that, at first order, we must-have for the dimensionless coupling $\tilde g_1$:
\begin{equation}
\frac{d \tilde g_1}{d \ln \Lambda} =-d_1 \tilde g_1+\mathcal{O}( \tilde g_1^2)\,. \label{eqflow1}
\end{equation}
The terms of order $\tilde g_1^2$ include the effects of the fluctuations, which are progressively integrated out in the RG procedure. To summarize, integrating fluctuations to each step change the fundamental scale $\Lambda\to \Lambda^\prime$ as well as the hamiltonian, and therefore the couplings constants, whose a part of the global modification comes from their proper dimension (providing by the term $-d_1 \tilde g_1$ in equation \eqref{eqflow1}). We will return on the exact computation of $d_1$ in the next section, in which we will present a nonperturbative formalism allowing us to compute non-Gaussian fixed points far from the Gaussian region. A practical limitation of the truncation \eqref{truncation1} comes from the choice of the representation. It is easier to do computations in the momentum space; however, translating the interaction in momentum space without knowledge of the eigenvectors leads to the mysterious quartic coupling:
\begin{equation}
\frac{g_1}{4!}\, \sum_i \phi_i^4=\frac{g_1}{4!}\,\sum_{\{\mu_j\}}\left( \sum_i \prod_{j=1}^4 u_i^{(\mu_j)}\right) \prod_{j=1}^4 \psi_{\mu_j}\,. \label{complic}
\end{equation}
A naive way to circumvent the difficulty is to note that, after all, the initial interaction is not fundamental. There are no experimental data to justify this interaction, and we may construct an approximation directly in momentum space. The original interaction is reminiscent to the familiar $\phi^4$ interaction $\int \phi^4(x) dx$, restricted to the positive (or negative) region. Heuristically, if we discard the boundary problems, the fields can be decomposed over ‘‘$\sin$" or ‘‘$\cos$" functions instead of ordinary Fourier transform; which are together eigenfunction for the ordinary Laplacian function. The coupling tensor in the bracket in the equation \eqref{complic} then behaves like:
\begin{equation}
\int \cos(k_1x)\cos(k_2x) \cos(k_3 x) \cos(k_4x) dx\,,
\end{equation}
which is essentially a sum of deltas of the form $\delta(k_1+\epsilon_2 k_2+\epsilon_3 k_3+\epsilon_4 k_4)$, with $\epsilon_i=\pm 1$. The symmetry of the delta function leads to two distinct couplings, following we have one or two positives $\epsilon_i$. By direct inspiration, we choose the following combination of Kronecker deltas:
\begin{equation}
\delta_1\mathcal{H} \sim \sum_{\{\mu_i\}} \, \left(\tilde g_1\delta_{0,p_1^2+p_2^2-p_3^2-p_4^2}+\tilde g_2\delta_{0,p_1^2+p_2^2+p_3^2-p_4^2}\right) \prod_{j=1}^4 \psi_{\mu_j}\,. \label{Hamiltonianessai}
\end{equation}
We expect that this Hamiltonian must have the same physical content than the original one \eqref{truncation1}. In particular, and even if it is a little bit caricatural, we expect that ensuring momentum conservation at the vertex level provides a well representation of the original local interaction $\sum_i \phi^4_i$. Unfortunately, the Hamiltonian \eqref{Hamiltonianessai} introduces a spurious singular behaviors at the origin for some one loop corrections. From a direct inspection, the problem come from the positivity of the momenta. A simple way to circumvent this difficulty is the following. Instead of positive eigenvalues, we consider the momentum $p$ as a relative integer $p\in \mathbb{Z}_{\sqrt{\Lambda-\lambda_0}}$; so that $p^2$ remains distributed following $\rho(\lambda)$, with $\lambda=p^2+\lambda_0$. Moreover, we introduce a new field $\psi(p)$, to distinguish them from $\psi_\mu$ considered above. Then, we chose for $\psi(p)$ the new Hamiltonian:
\begin{align}
\nonumber \mathcal{H}[\Psi]=\frac{1}{2} \,\sum_p & \, \psi(-p)(p^2+\lambda_0)\psi(p)\\
\qquad+& \frac{\tilde g}{4!} \sum_{\{p_i\}} \,\delta_{0,p_1+p_2+p_3+p_4} \prod_{j=1}^4 \psi(p_j)\,, \label{themodel}
\end{align}
To summarize, from the Hamiltonian \eqref{truncation1}, we kept essentially three elements:
\begin{enumerate}
\item Without interaction, the correlation functions are essentially given by the eigenvalues of the covariance matrix $\mathcal{C}$.
\item The square of the momenta $p^2$ are distributed following the distribution $\rho(\lambda)$.
\item The interactions are essentially locals in the usual sense.
\end{enumerate}
We expect that for our investigations about the stability of the Gaussian distribution, only these three points are really relevant; and we only consider the Hamiltonian \eqref{themodel} for explicit computations using nonperturbative formalism in section \ref{sec3} and \ref{sec4}. However, we will continue to use the Hamiltonian \eqref{truncation1} as well for some discussion, to make contact with the reference \cite{pca0}. Note to conclude that, in contrast to the coupling $g$ in \eqref{truncation1}, the coupling $\tilde{g}$ involved in \eqref{themodel} have to be of order $1/N$ in order to ensure extensivity of the model. This can be checked as follow. In the words of physicists, let us consider a “one loop correction" to the $2$-point function (i.e. a correction of order $g$):
\begin{equation}
\langle \phi_i \phi_j \rangle \sim \mathcal{C}_{ii} \delta_{ij}\,,
\end{equation}
then; setting $i=j$ and summing over $i$:
\begin{equation}
\sum_{i}\langle \phi_i \phi_i \rangle \sim \Tr(\mathcal{C})=\sum_\ell \lambda^{-1}_\ell\,.
\end{equation}
In contrast, let us consider the same kind of correction for the Hamiltonian \eqref{themodel},
\begin{equation}
\langle \psi_{\mu} \psi_{\mu^\prime}\rangle \sim \delta_{p_\mu,p_{\mu^\prime}} \sum_\ell \lambda^{-1}_\ell\,,
\end{equation}
so, setting $p_\mu=p_{\mu^\prime}$ and summing over $p_\mu$, we get:
\begin{equation}
\sum_{\mu} \langle \psi_{\mu} \psi_{\mu}\rangle \sim \left(\sum_{\mu}1\right) \sum_\ell \lambda^{-1}_\ell= N\,\sum_\ell \lambda^{-1}_\ell\,.
\end{equation}
Therefore, to ensure extensivity, the coupling with tide have to be of order $1/N$, as expected. To make this dependence explicit, we keep the following expression, without tilde:
\begin{align}
\nonumber \mathcal{H}[\Psi]=\frac{1}{2} \,\sum_p & \, \psi(-p)(p^2+\lambda_0)\psi(p)\\
\qquad+& \frac{g}{4!N} \sum_{\{p_i\}} \,\delta_{0,p_1+p_2+p_3+p_4} \prod_{j=1}^4 \psi(p_j)\,. \label{themodel2}
\end{align}
We will return on this scaling at the moment of the derivation of the flow equations, in section \ref{sec4} below.

\section{A nonperturbative framework} \label{sec3}

\subsection{The exact RG equation}

One expects that the accuracy of the results obtained in \cite{pca0} may be improved taking into account higher couplings and loop effects, motivating a nonperturbative analysis. The most powerful formalism to keep the nonperturbative effect of the RG is the functional renormalization group (FRG) formalism, essentially based on the Wetterich-Morris equation \cite{Wilson:1971dc}-\cite{Wetterich:1992yh}. In this section, we propose to construct a version of this formalism adapted to the PCA investigations. As we recalled, the Wilson-Kadanoff procedure requires splitting into modes, between UV scales (no fluctuations are integrated out) and IR scales (all the fluctuations are integrated out) dictating how the small distance fluctuations are integrated out. In the FRG formalism, this progressive integration of UV modes work thinks to a \textit{momentum-dependent mass} term $\Delta \mathcal{H}_k[\Phi]$ added to the microscopic hamiltonian $\mathcal{H}[\Phi]$. In momentum representation:
\begin{equation}
\Delta \mathcal{H}_k[\Psi]=\frac{1}{2}\sum_{p} \psi(-p) \, r_k(p^2) \,\psi(p)\,,
\end{equation}
where $k$ play the role of a referent momentum scale. The regulator function $r_k(p^2) $ have to satisfy some elementary requirements, in such a way that:
\begin{enumerate}
\item Small distance fluctuations ($p^2 >k^2$) are unaffected by the presence of $\Delta \mathcal{H}_k[\Psi]$ and integrated out.
\item Long distance fluctuations ($p^2 <k^2$) acquire a large mass and are frozen out.
\end{enumerate}
In this way, the momentum-dependent mass $r_k(p^2)$ must be satisfy the elementary requirements:
\begin{enumerate}
\item $r_k(p^2)$ has to have a non-vanishing infrared limit, $p^2/k^2\to 0$\,,
\item $r_k(p^2)\to 0$ in the ultraviolet limit, for $p^2/k^2\gg 1$\,,
\item $r_k(p^2)$ has to vanish in the limit $k\to 0$, allowing to recover the original partition function when all the degrees of freedom are integrated out\,,
\item $r_k(p^2)$ has to be of order $\Lambda$ for $k^2\to \Lambda$, $\Lambda$ referring to the larger eigenvalue of $\mathcal{K}$.
\end{enumerate}
Introducing this mass term into the microscopic hamiltonian, we replace the global description given by the initial generating functional $\mathcal{Z}[J]:=\int d\Psi e^{-\mathcal{H}[\Psi]+ \sum_p j(p)\psi(p)}$, by a one-parameter set of models indexed by $k$, $\{\mathcal{Z}_k[J]\}$ defined as:
\begin{equation}
\mathcal{Z}_k[J]:=\int\, d\Psi\, e^{-\mathcal{H}[\Psi]-\Delta \mathcal{H}_k [\Psi]+\sum_p j(p)\psi(p)}\label{part2}\,,
\end{equation}
When the scale $k$ decrease, more and more degrees of freedom are integrated out. The infinitesimal transcription of this goes through a first order differential equation \cite{Wilson:1971dc}-\cite{Wetterich:1992yh}:
\begin{align}
\nonumber \dot{\Gamma}_k&=\frac{1}{2} \sum_\mu\, \dot{r}_k(p_\mu^2)\left(\Gamma^{(2)}_k+r_k\right)^{-1}_{\mu,-\mu}\\
&=N \int_{p\geq 0} dp\,\tilde{\rho}(p^2)p \dot{r}_k(p^2)\left(\Gamma^{(2)}_k+r_k\right)^{-1}(p,-p) \,,\label{Wett}
\end{align}
where to write the last line we introduced the momentum density $ \tilde{\rho}(p^2)$, related to the eigenvalues density ${\rho}(\lambda)$ as:
\begin{equation}
{\tilde{\rho}(p^2)} := {\rho}(p^2+\lambda_0):=\frac{1}{N} \sum_\mu \delta(\lambda-\lambda_\mu)\,.
\end{equation}
Note that the normalization has to be chosen such that the number of degree of freedom remains equal to $N$:
\begin{equation}
N \int_{p\geq 0} \tilde{\rho}(p^2) pdp= \frac{N}{2} \, \int d\lambda \rho(\lambda)=\frac{N}{2}\,.
\end{equation}
Equation \eqref{Wett} indicates how the \textit{average effective hamiltonian} $\Gamma_k$ change in the windows of scale $[k-dk,k]$ -- the dot meaning derivative with respect to the RG parameter $t:=\ln k$: $\dot{X}=k\frac{d}{dk}X$. We recall that the average effective action is defined as slightly modified Legendre transform of the free energy $\mathcal{W}_k:=\ln\mathcal{Z}_k$ :
\begin{equation}
\Gamma_k[M]+\Delta \mathcal{H}_k[M]=\sum_p j(p) m(p)-\mathcal{W}_k[J]\,,
\end{equation}
with $M:=\{m(p)\}$. Moreover, note that $\Gamma^{(2)}_k$ in equation \eqref{Wett} denotes the second derivative of the average effective action :
\begin{equation}
\left[\Gamma^{(2)}_{k}\right]_{\mu\mu^\prime}:=\frac{\partial^2 \Gamma_k}{\partial m(p_{\mu})\partial m(p_{\mu^\prime})}\,,
\end{equation}
where $m(p_\mu)$ is defined as: $m(p)=\partial\mathcal{W}_k[J]/\partial j(p) $.

\subsection{ Scaling dimension}
\begin{figure}
\includegraphics[scale=0.6]{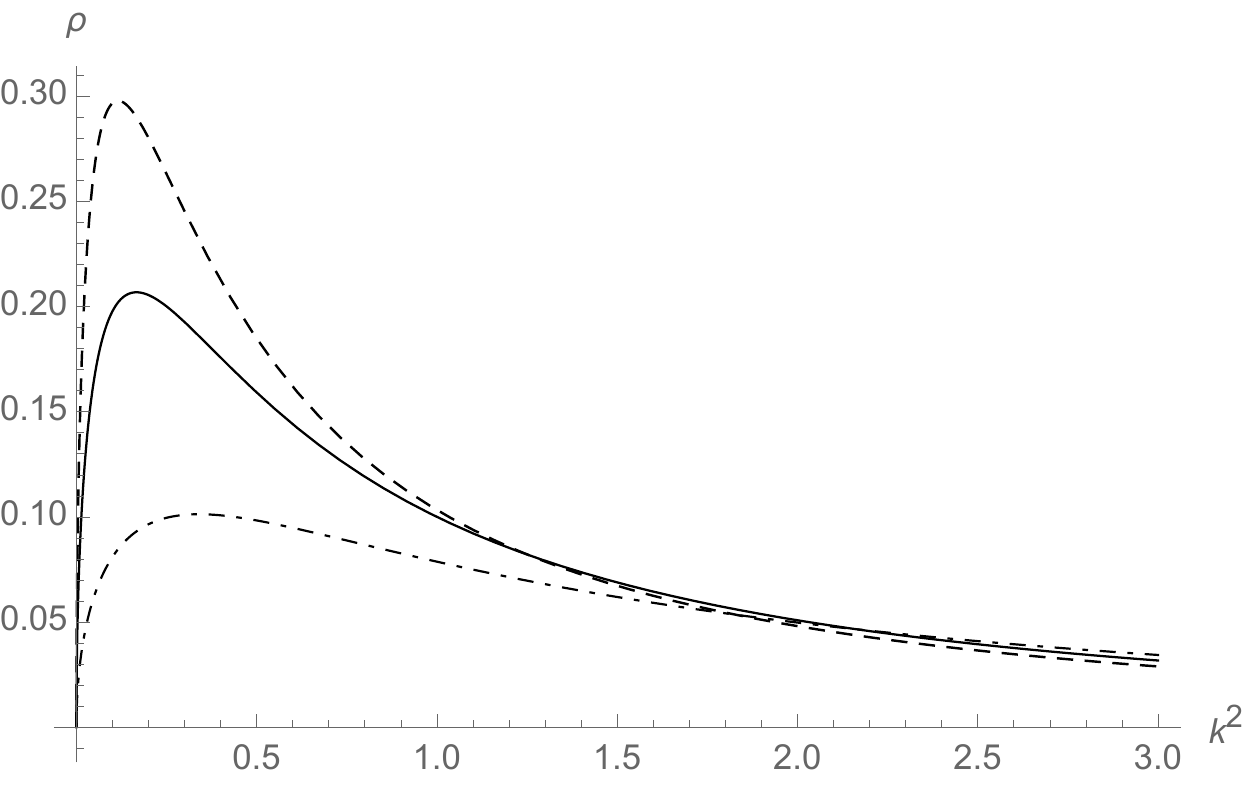}
\caption{The MP momentum representation of the MP distribution in the deep UV. The dashed curve is for $\sigma^2=1.2$, the solid curve for $\sigma^2=1$ and the dashed-dotted curve for $\sigma^2=0.7$. } \label{MP1}
\end{figure}

 The scaling (or canonical) dimension is defined as the intrinsic dependence of a quantity on the cut-off coming from its dimension. In standard quantum field theory, for instance, the dimensions of the couplings are closely related to renormalizability. In this case, the dimensions are inherited from the referent background space-time where the field is described. The difficulty here comes from that we do not have any background space-time in \eqref{part2} to fix the dimensions. However, it may be instructive to return to the standard field theory case. Let us for instance consider a free scalar field $\varphi(x)$ defined over $\mathbb{R}^d$; with hamiltonian:
\begin{equation}
\mathcal{H}=\int dx\, \varphi(x)(-\bigtriangleup)\varphi(x)\,.
\end{equation}
Here, the role of the matrix $\mathcal{K}$ is played by the Laplacian $\bigtriangleup$, whose spectrum fix the size of the fluctuations and discriminate between IR and UV scales. The eigenvalues will be denoted as $p^2$, referring to their positivity, and we must have $[p]=-[x]$; $[X]$ denoting the dimension of the quantity $X$. We fix the definition of the bracket such that $[p]=1$. The dimension of the fields $\varphi(x)$ can be read directly from the previous expression from the requirement that $\mathcal{H}$ have to be dimensionless, we get:
\begin{equation}
[\varphi]=\frac{d-2}{2}\,. \label{dimfield}
\end{equation}
Now, let us consider the $2$-point function $\langle \varphi(x) \varphi(x)\rangle$,
\begin{equation}
\langle \varphi(x) \varphi(x)\rangle \sim \frac{1}{p^2}\,,
\end{equation}
so that:
\begin{equation}
\int dx\,\langle \varphi(x) \varphi(x)\rangle \sim \int^\Lambda \, \frac{dp}{p^2} \propto \Lambda^{d-2}
\end{equation}
for some UV cut-off $\Lambda$. The dependence on $\Lambda$ reflect the dimension of the field given by \eqref{dimfield}; and suggest a way to define dimension without referent background. It must be fixed by the rescaling $\varphi\to \bar\varphi = \Lambda^{-[\varphi]}\varphi$ such that $ \int dx\,\langle \bar\varphi(x) \bar\varphi(x)\rangle $ becomes essentially independent of $\Lambda$ for sufficiently large $\Lambda$:
\begin{equation}
\frac{d}{d\Lambda} \,\int dx\,\langle \bar\varphi(x) \bar\varphi(x)\rangle \approx 0\,.
\end{equation}
By direct inspiration, we fix the rescaling $z_\Lambda$ of the variables $\phi_i$, $\phi_i\to \tilde{\phi}_i:=z_\Lambda \phi_i$ such that $\sum_i \langle \tilde{\phi}_i^2 \rangle$ becomes $\Lambda$-independent:
\begin{equation}
\frac{d}{d\Lambda} \sum_i \langle (z_\Lambda \phi_i)^2\rangle =0 \to \frac{d \ln z_\Lambda}{d\ln\Lambda}=- \frac{1}{2} \frac{\rho(\Lambda)}{\int d\lambda \frac{\rho(\lambda)}{\lambda}} \,.
\end{equation}
The dimension of the coupling constant must be fixed following the same strategy. The field dimension $d_1=-\ln z_\Lambda$ being fixed, the dimension of the quartic coupling $g$ must be fixed from the extensivity argument used in \cite{pca0}. The authors argue that the average of the hamiltonian (which is essentially the energy in physics) must be an extensive quantity, and therefore proportional to the number of effective degrees of freedom. This requirement in particular ensures that the entropy is an extensive quantity as well. For some cut-off $\Lambda$, the number of effective degrees of freedom $N_{\text{eff}}$ is nothing but the number of eigenvalues $\lambda_\mu$, then we must have:
\begin{equation}
N_{\text{eff}}:=N\int^\Lambda d\lambda\, \rho(\lambda)\,.
\end{equation}
Therefore, one expects that the rescaling allowing to pass from $g_1$ to the dimensionless coupling $\bar{g}_1$ must be such that:
\begin{equation}
N g_1 \frac{1}{N}\sum_i \phi_i^4 = N_{\text{eff}}\, \bar{g}_1 \, \sum_i \frac{1}{N}\,(z_\Lambda \phi_i)^4\,, \label{dim1}
\end{equation}
which fix the rescaling of the coupling constant as:
\begin{equation}
\frac{d\ln \bar{g}_1}{d\ln \Lambda}= \rho(\Lambda) \left[\frac{2}{\int d\lambda \frac{\rho(\lambda)}{\lambda}} - \frac{\Lambda}{\int d\lambda \rho(\lambda)} \right]\,.
\end{equation}
The formula can be easily generalized for an interaction involving $p$ fields;
\begin{equation}
\delta \mathcal{H}\propto g_p \sum_{i_1,\cdots,i_p} \, \mathcal{V}_{i_1,\cdots, i_p} \prod_{a=1}^{2p} \phi_{i_a}\equiv g_p \mathcal{V}[\phi^{2p}]\,,
\end{equation}
where the symbol $\mathcal{V}_{i_1,\cdots, i_p}$ must be a product of Kronecker delta, identifying indices pairwise. Let us denote as $n(\mathcal{V})$ the number of Kronecker delta. Therefore, equation \eqref{dim1} have to be replaced by:
\begin{align}
N g_p \frac{1}{N^{p-n(\mathcal{V})}}&\mathcal{V}[\phi^{2p}]= N_{\text{eff}} \,z_{\Lambda}^{2p}\, \bar{g}_p \,\frac{1}{N^{p-n(\mathcal{V})}} \mathcal{V}[\phi^{2p}]
\end{align}
leading to:
\begin{equation}
\frac{d\ln \bar{g}_p}{d\ln \Lambda}= \rho(\Lambda) \left[\frac{p}{\int d\lambda \frac{\rho(\lambda)}{\lambda}} - \frac{\Lambda}{\int d\lambda \rho(\lambda)} \right]\,. \label{dim2}
\end{equation}
The canonical dimension that we discussed here corresponds to the one discussed in \cite{pca0}. However, for FRG applications, we must have to find the scaling concerning the running scale $k$, not concerning the fundamental cut-off $\Lambda$. Once again, the question is trivial for standard quantum field theory, the dimension being the same that we use $\Lambda$ of $k$ as referent scale. But here, this is not trivial, because the definition of the canonical dimension seems to introduce a $\Lambda$ dependence for some distribution $\rho(\lambda)$. In contrast, the distribution $\dot{r}_k$ ensures that only windows of momenta around $p_\mu = k$ contribute significantly to the integral in the right-hand side of \eqref{Wett}. This is why it more natural to define the canonical scaling for the running scale, and fix the dimensions concerning this parameter, rather than for $\Lambda$, which introduce a spurious reference to the microscopic physics. This difficulty may be solved in the same way as we defined the canonical dimension for fields. Returning on the free scalar field $\varphi(x)$, for some regulator $r_k(p^2)$ in Fourier space, we must have:
\begin{equation*}
\int dxdy\,\langle \varphi(x)\dot{r}_k(x-y) \varphi(y)\rangle \sim \int^\Lambda \, \frac{dp}{p^2}\dot{r}(p^2) \propto k^{d-2+[{r}_k]} \,.
\end{equation*}
Therefore it must be possible to use this relation to define the dimension; as the rescaling of the fields such that the right hand side scale as $k^{[{r}_k]}$. Moving on to our field theory for $\phi_i$, it is clear from definition of $r_k$ that $[{r}_k]$ must be equal to $1$. Then, one expect that there exist a rescaling $z_k$ of the fields, such that:
\begin{equation}
\frac{d}{dk} \, \sum_{i,j} \langle (z_k\phi_i)(\dot{r}_kr_k^{-1})_{ij}(z_k\phi_j)\rangle \approx 0\,,
\end{equation}
for sufficiently large $k$. The two very definitions, for $z_\Lambda$ and $z_k$ seems to be different. However, a moment of reflection show that they have to coincide at least in the deep UV\footnote{Note that the dimensions may be fixed from a purely RG point of view, from the behavior of the RG flow in the vicinity of the Gaussian fixed point. More precisely, requiring that the first leading order perturbative corrections have the same scaling as the corresponding couplings provides a notion of dimension, which reduce to the previous one especially for power-law distributions $\rho(\lambda)\propto \lambda^{\alpha}$; provided that $\alpha>-1$. For instance, assuming that $\lambda_0$ has dimension $1$ in $\Lambda$, we find that the first leading order quantum correction to mass scales as $\Lambda^{\alpha}$, and therefore requires that the corresponding coupling scale as $\Lambda^{1-\alpha}$ in order to get a global scaling identical to the first term. Which is nothing but that we get explicitly from equation \eqref{dim2}. We will return on this approach in the next section.}, for $k$ and $\Lambda$ very larges. In fact, it is easy to cheek that for power-law distributions $\rho(\lambda)\propto \lambda^{\alpha}$, the right hand sides of equations \eqref{dim1} and \eqref{dim2} becomes pure numbers depending only on $\alpha$; which is a general feature of homogeneous distributions $\tilde\rho(a p)=a^\beta \tilde\rho(p)$. For all cases, it is easy to cheek that the dimensions are the same using the momentum cut-off $\Lambda$ and the distribution $\dot{r}_k$, for sufficiently large $k$. In the next sections, we will compare the flow obtained from spectra with signal and purely noisy signals. The Marchenko-Pastur (MP) distribution usually provides a well efficient description of noisy signal; and it corresponds to the asymptotic spectrum of purely i.i.d random matrices $X_{ai}$, with arbitrary larges $p$ and $n$, but $p/n$ kept constant \cite{Kanzieper:2010xr}-\cite{Lu:2014jua}. Explicitly, the eigenvalue distribution $\mu(x)$ is the following
\begin{equation}
\mu(x)=\frac{1}{2\pi \sigma^2} \frac{\sqrt{(a_+-x)(x-a_-)}}{kx}\,,\label{MPL}
\end{equation}
where :
\begin{itemize}
\item $k=p/n$ is the fixed ratio between the size indices of the random matrix $X_{ai}$ with i.i.d entries, with $0$ means and variance $\sigma < \infty $.
\item $a_{\pm}=\sigma^2(1\pm \sqrt{k})^2$.
\end{itemize}
Now, let us consider the behavior of the distribution for large $x$, i.e. for $x$ close to $a_+$ (the maximal eigenvalue); and expand it in power of $t=x-b$. At leading order, we get straightforwardly in the UV regime $1/a_{-}\gg 1/\lambda $:
\begin{equation}
\tilde{\rho}(p^2)\approx \frac{1+\sqrt{k}}{2\pi\sigma k} \frac{(p^2)^{1/2}}{(p^2+m^2)^2} \,, \label{MPscaling}
\end{equation}
where $m^2:=1/a_+$. The corresponding distribution is plotted on Figure \eqref{MP1}. Note that the distribution \eqref{MPL} is for the inverse of the kinetic kernel $\mathcal{K}=\mathcal{C}^{-1}$; in contrast with the law \eqref{MPscaling}, as the notations suggest. The relation between $\mu$ and $\rho$ could be easily deduced from the fact the number of eigenvalues is the same for $\mathcal{K}$ and $\mathcal{C}$, leading to:
\begin{equation}
\rho(x)=\mu(x^{-1})\frac{1}{x^2}\,.
\end{equation}

\section{Solving RG using local potential approximation} \label{sec4}
Solving the exact nonperturbative RG flow equation \eqref{Wett} is a difficult task, even in very special cases. Therefore, extracting some information about this equation requires approximations.
The difficulty to solve the exact RG equation \eqref{Wett} may be pointed out as follow. Taking the second derivative of \eqref{Wett} with respect to the classical field $m_\mu$; $\partial^2/\partial m_\mu\partial m_{\mu^\prime}$, we get an equation for $\dot{\Gamma}_k^{(2)}$. Assuming that odd functions $\Gamma^{(2n+1)}_k$ vanish identically, we get that the right hand side involve $\Gamma^{(4)}_k$ and the effective propagator $G_k:=(\Gamma^{(2)}_k+r_k)^{-1}$. Deriving once again two times concerning the classical field, we get an equation for $\dot{\Gamma}_k^{(4)}$, involving $G_k$, ${\Gamma}_k^{(4)}$ and ${\Gamma}_k^{(6)}$, and so on. Taking successive derivatives, we then generate an infinite tower of coupled equations. All the approximation schemes used to solve the RG equations have to aim to close this hierarchy. In this paper we focus on the crude truncation approximation, imposing:
\begin{equation}
\Gamma_k^{(2n)}\approx 0\,,
\end{equation}
until a certain $n$. The restriction to even functions, i.e. $\Gamma^{(2n+1)}_k=0$ reflect the $\phi\to -\phi$ symmetry of the original microscopic action, and corresponds to expands the truncated action $\Gamma_k$ around vanishing classical field $m_\mu=0$, $\forall \mu$. We call \textit{symmetric phase} the portion of the phase space parametrized like that; and in this introductory paper, we only focus on this approximation. Focusing on this approximation, we consider the following truncation around $n=3$:
\begin{align}
\nonumber \Gamma_k[M]=\frac{1}{2} \,\sum_p\, &m(-p)(p^2+u_2(k))m(p)
\\
&+ \sum_{\{p_i\}} \, \bigg(\frac{g({k})}{4!N}\delta_{0,\sum_ip_i}\bigg) \prod_{j=1}^4 m(p_j)\,. \label{thetruncation}
\end{align}
Such a truncation define a parametrization of the \textit{theory space}, i.e. the space of allowed actions, and is called local potential approximation (LPA) \cite{Delamotte:2007pf}-\cite{Pawlowski:2015mlf}; the momentum dependence on the effective vertex being completely discarded. One expects that such an approximation work well for this kind of theories \cite{Delamotte:2007pf}. The dependence on $k$ for the coupling constant $g({k})$ and for effective mass $u_2(k)$ reflect the integration of UV degrees of freedom when $k$ varies on equation \eqref{Wett}. Starting calculations still requires two ingredients: the momentum distribution and the regulator function $r_k$. For the last one, we choose the standard Litim regulator \cite{Litim:2000ci}-\cite{Litim:2001dt}, allowing to do analytic computations:
\begin{equation}
r_k(p^2)=(k^2-p^2)\theta(k^2-p^2)\,, \label{litim}
\end{equation}
the $\theta(x)$ being the Heaviside step function, equal to $1$ for $x\geq 1$ and to $0$ otherwise. Note that physical solutions of the exact RG flow equation \eqref{Wett}, in principle, do not depend on $r_k$. However, the approximation used to solve it generally introduce a dependence on the regulator, which have to be investigated. The regulator \eqref{litim} has been investigated to be optimal for such dependence for some models, see \cite{Canet:2002gs}, which is another practical advantage to do this choice. The remaining ingredient is the momentum distribution. In principle, this distribution has to come from a data set and has not analytic form as for the MP law. To keep contact with the reference paper \cite{pca0}; we will firstly consider the example of a power lay distribution $\tilde{\rho}(p^2)\propto p^{2\alpha}$. Note that ordinary field theories provide a non-trivial example of such a power-law behavior, the value of $\alpha$ being trivially connected to the space-dimension through $\alpha=D/2-1$. However, for spectra relevant in data analysis, such an approximation cannot be considered better than a caricature or a limit case. For instance, for $p\ll 1$, the MP law \eqref{MPscaling} is such that $\tilde{\rho}(p^2)\propto (p^{2})^{1/2}$; but remains true only in the deep IR sector. However, to investigate the behavior of the RG flow for such a power-law remains an instructive exercise for the unfamiliar reader, to show how the different concepts like canonical dimension work, and to show how this behavior is influenced by the shape of the distribution. We thus briefly discuss this case in the next section, before moving on to more realistic spectra.

\subsection{Power law distribution}\label{powerlaw}
\noindent
The flow equations for $\dot{g}$, and $\dot{u}_2$ can be deduced by projection of the exact flow equation \eqref{Wett} along the reduced portion of the full phase space parametrized by the truncation \eqref{thetruncation}. To implement this, let us consider the derivative of \eqref{Wett} with respect to $m(p)$ and $m(p^\prime)$. Because $\Gamma^{(3)}_k=0$; we get:
\begin{equation}
\dot{\Gamma}_{k,\mu_1\mu_2}^{(2)}=- \frac{1}{2} \sum_\mu \dot{r}_k(p_\mu^2) G_{k,\mu\mu^\prime} \Gamma^{(4)}_{k,\mu^\prime\mu^{\prime\prime}\mu_1\mu_2} G_{k,\mu^{\prime\prime}\mu}\,.\label{exp1}
\end{equation}
Setting $\mu_1=\mu_2$, from the truncation \eqref{thetruncation}, it follows that:
\begin{equation}
\dot{\Gamma}_{k,\mu_1\mu_1}^{(2)}=\dot{u}_2\,.
\end{equation}
Moreover, the derivative of ${r}_k(p_\mu)$ can be easily computed: $\dot{r}_k(p^2)=2k^2\theta(k^2-p^2)$. Finally, in the symmetric phase, $\Gamma^{(2)}_k$ must be easily computed:
\begin{equation}
\Gamma^{(2)}_{k,\mu_1\mu_2}=\delta_{p_{\mu_1},-p_{\mu_2}} \left(p_{\mu_1}^2+u_2(k)\right)\,,
\end{equation}
and \eqref{exp1} reduces to:
\begin{equation}
\dot{u}_2= -\frac{1}{2}\frac{2k^2}{(k^2+u_2)^2} \sum_\mu \theta(k^2-p^2_\mu) \Gamma^{(4)}_{k,\mu\mu \mu_1\mu_1}\bigg\vert_{p_{\mu_1}=0}\,.\label{flow}
\end{equation}
The last sum involves sums over different permutations of external momenta, arising from derivations. However, for large $k$, all these terms do not provide a significant contribution. Let us consider the contribution arising from the coupling $g$, we have:
\begin{equation}
\Gamma^{(4)}_{k,\mu_1\mu_2 \mu_3\mu_4}=\frac{g}{4!N}\sum_\pi \delta_{0,p_{\pi(1)}+p_{\pi(2)}+p_{\pi(3)}+p_{\pi(4)}}\,,\label{perm}
\end{equation}
the sum running over the set of permutation of the four external momenta. Because all the momenta play the same role, all the permutations contribute to the right hand side of the equation \eqref{flow}, the typical contribution writing as:
\begin{equation}
\frac{1}{N}\sum_\mu \theta(k^2-p^2_\mu) \frac{g}{4!} \delta_{0,p_1+p_2} \sim k^{2\alpha+2} \,. \label{flow2prime}
\end{equation}
Therefore, keeping into account the scaling of the coupling constant, we get a global $k$ dependence as $k^{2\alpha+2} k^2 k^{-4} k^{2-2\alpha} = k^2$, which is nothing but what we expect for the proper scaling of $u_2$. Note that, as explained in footnote 4 at the end of section \ref{sec3}, the scaling may be deduced directly from these equations. Indeed, assuming that $g$ scales as $k^{d_1}$, we get that contributions as \eqref{flow2prime} scales as $k^{d_1-2+2(\alpha+1)}$. Moreover, the scaling of $u_2$ which is essentially the lower eigenvalue of the spectrum of the kinetic kernel -- must scale as $\Lambda$, and therefore as\footnote{All the eigenvalues must have to be homogeneous to the upper eigenvalue, ensuring that under a global dilatation the shape of the spectrum remains unchanged.} $k^2$. Then, we must have $d_1=2(1-\alpha)$. 
\medskip

\begin{figure}
 \includegraphics[scale=0.32]{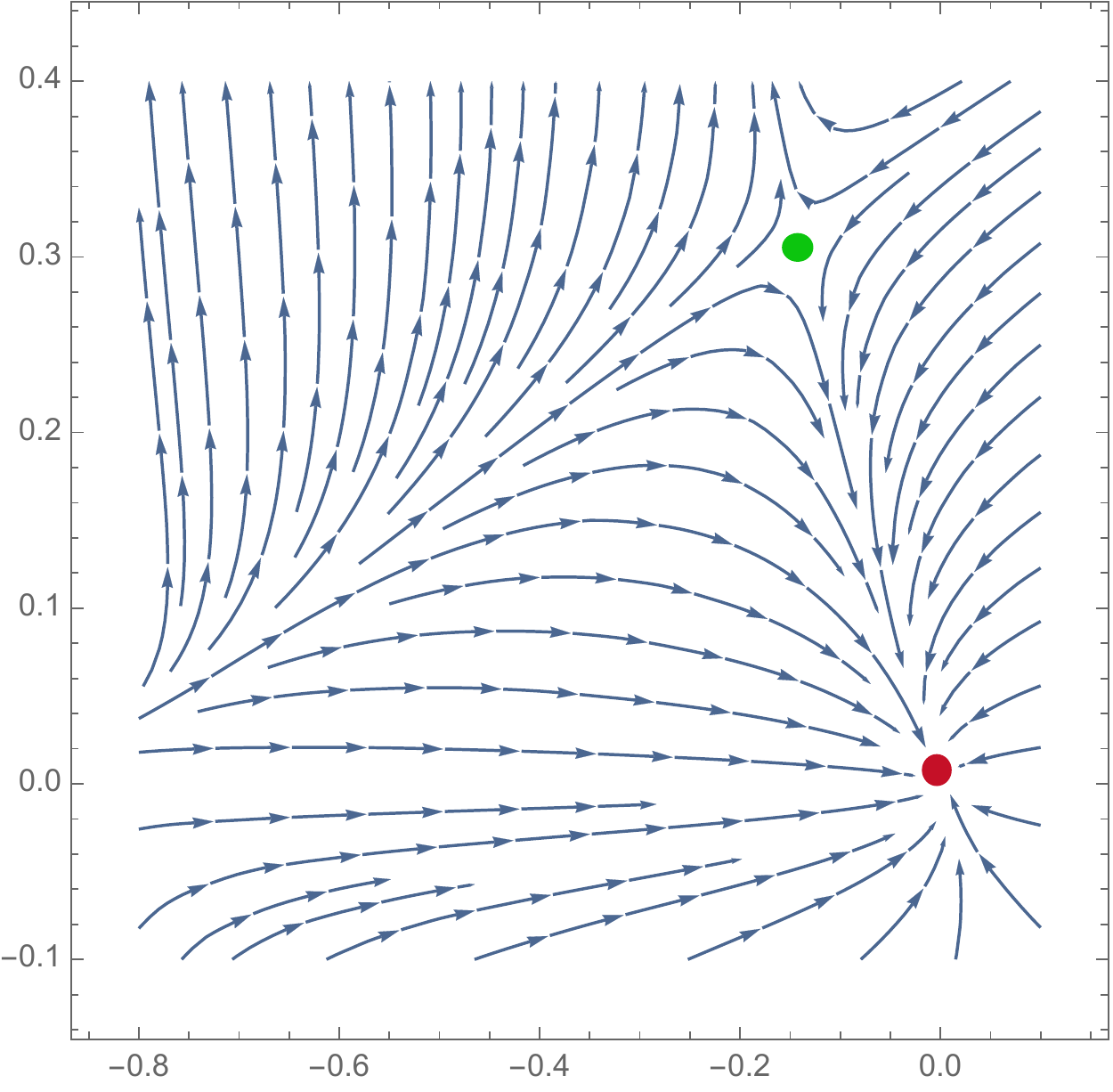}\,\includegraphics[scale=0.32]{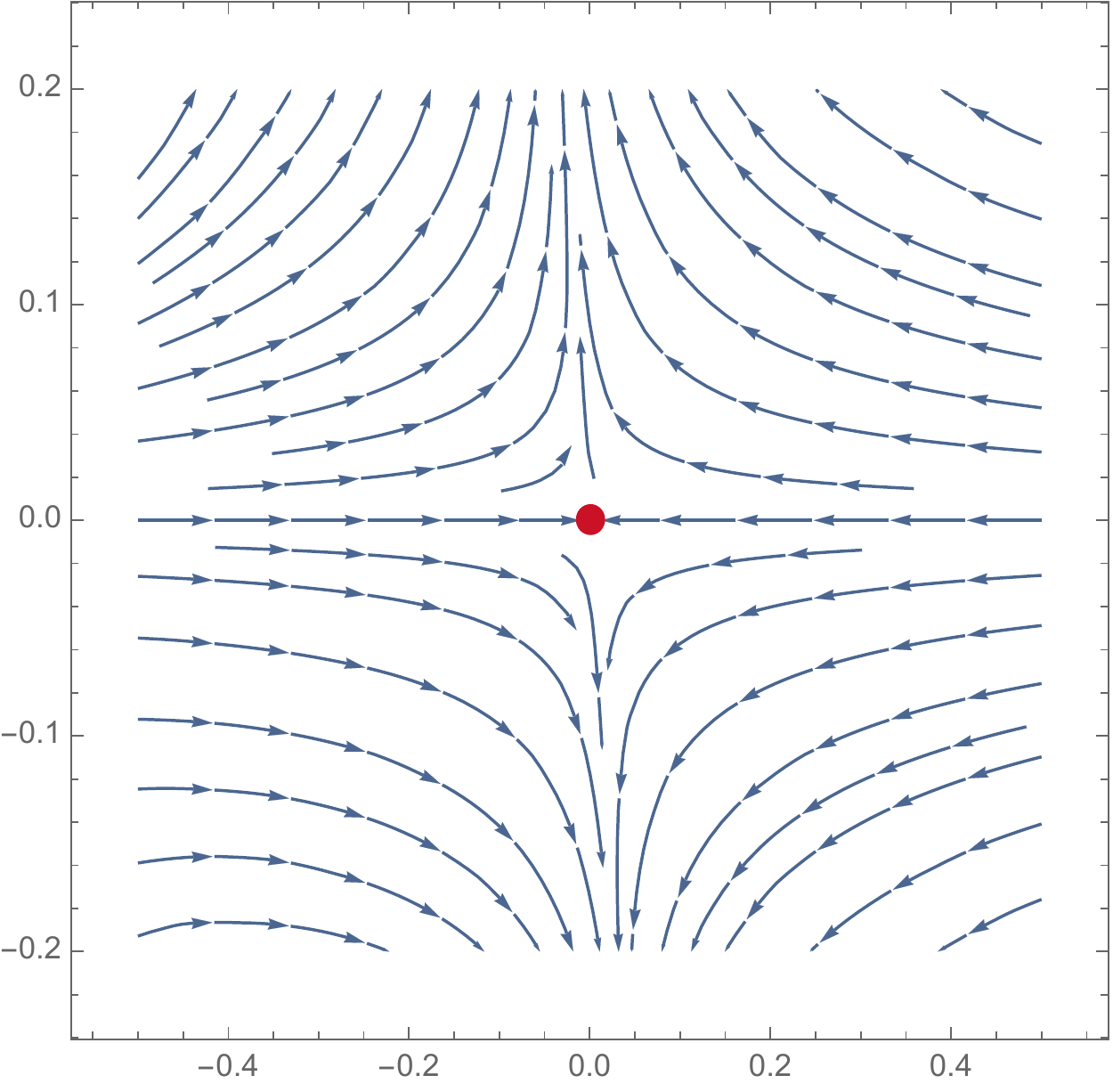}
\caption{From left to right, numerical RG flow for $\alpha=0.5$ and $\alpha=1.5$. In both cases the red and green points correspond respectively to Gaussian and non-Gaussian fixed points, and the arrows are oriented toward UV scales (from small to big $k$). } \label{fig3}
\end{figure}

From these observations, and counting the number of relevant contractions in \eqref{flow}, we obtain:
\begin{equation}
\dot{u}_2= -\dfrac{g}{1+\alpha}\dfrac{k^{4+2\alpha}}{(k^2+u_2)^2} \,.\label{flowbert}
\end{equation}
This equation involves couplings having dimensions. Due to the fact that it is suitable to work with dimensionless couplings, we introduce the dimensionless parameters:
\begin{equation}
u_2=: k^2 \bar{u}_2\,, \quad g =: k^{2(1-\alpha)} \bar{g}\,,
\end{equation}
leading to ($\beta_2:= \dot{\bar{u}}_2$):
\begin{equation}
\beta_2= -2\bar{u}_2-\dfrac{\bar{g}}{1+\alpha}\dfrac{1}{(1+\bar{u}_2)^2}\,. \label{beta1}
\end{equation}
To find the equation for $\dot{g}$, we proceed exactly on the same way. We take the fourth derivative of the flow equation \eqref{Wett}, applying $\partial^4/\partial m_{\mu_1}\partial m_{\mu_2}\partial m_{\mu_3}\partial m_{\mu_4}$, and setting all the external momenta to be equals,
\begin{align*}
\dot{\Gamma}_{k,\mu_1\mu_1\mu_1\mu_1}^{(4)}=&\,3\sum_{\mu,\mu^\prime,\mu^{\prime\prime},\mu^{\prime\prime\prime},\mu^{\prime\prime\prime\prime}} \dot{r}_k(\mu) G_{\mu\mu^\prime} \Gamma^{(4)}_{k,\mu^{\prime}\mu^{\prime\prime}\mu_1\mu_1}\\
& \qquad\quad \times G_{\mu^{\prime\prime} \mu^{\prime\prime\prime} } \Gamma^{(4)}_{k,\mu^{\prime\prime\prime}\mu^{\prime\prime\prime\prime}\mu_1\mu_1} G_{\mu^{\prime\prime\prime\prime}\mu}\,.
\end{align*}
From the truncation, it follow that: $\dot{\Gamma}_{k,0,0,0,0}^{(4)}=\frac{\dot{g}}{N}$. Thus, repeating the same analysis as for the flow of $u_2$, we get for the dimensionless coupling $\bar{g}$ ($\beta_g:=\dot{\bar{g}}$):
\begin{equation}
\beta_g=-2(1-\alpha)\bar{g}+\frac{6}{1+\alpha}\,\frac{\bar{g}^2}{(1+\bar{u}_2)^3}\,. \label{beta2}
\end{equation}
The flow equations \eqref{beta1} and \eqref{beta2} exhibit fixed points, which can be easily found solving the system $\beta_2=\beta_4=0$. In addition to the Gaussian fixed point, with $\bar{g}=\bar{u}_2=0$, we get the non-Gaussian fixed point:
\begin{equation}
p=(\bar u_2^*, \bar{g}^*)=\left(-1-\frac{6}{\alpha-7},72\frac{ \alpha^2-1}{( \alpha-7)^3}\right)\,,
\end{equation}
which provides a explicit example of how the RG flow depends on $\alpha$. In particular, we see that the value $\alpha=1$ corresponds to what we call a critical dimension in standard field theory. For $\alpha<1$, a fixed point is  reminiscent of the standard Wilson-Fisher (WF) fixed point; with one attractive and one repulsive direction\footnote{We recall the standard vocabulary in physics: In the vicinity of a fixed point, a direction is relevant (toward the UV scales) if the eigenvalue is positive, irrelevant if it is negative and marginal if it is zero. Moreover, for the Gaussian fixed point, the eigenvalues are nothing but the canonical dimensions.}. For $\alpha>1$ however, the behavior of the RG flow is governed by the Gaussian fixed point, which becomes unstable (See Figures \ref{fig3} for an illustration).

\subsection{Flow equations without power law approximation}
\begin{figure}
\includegraphics[scale=0.55]{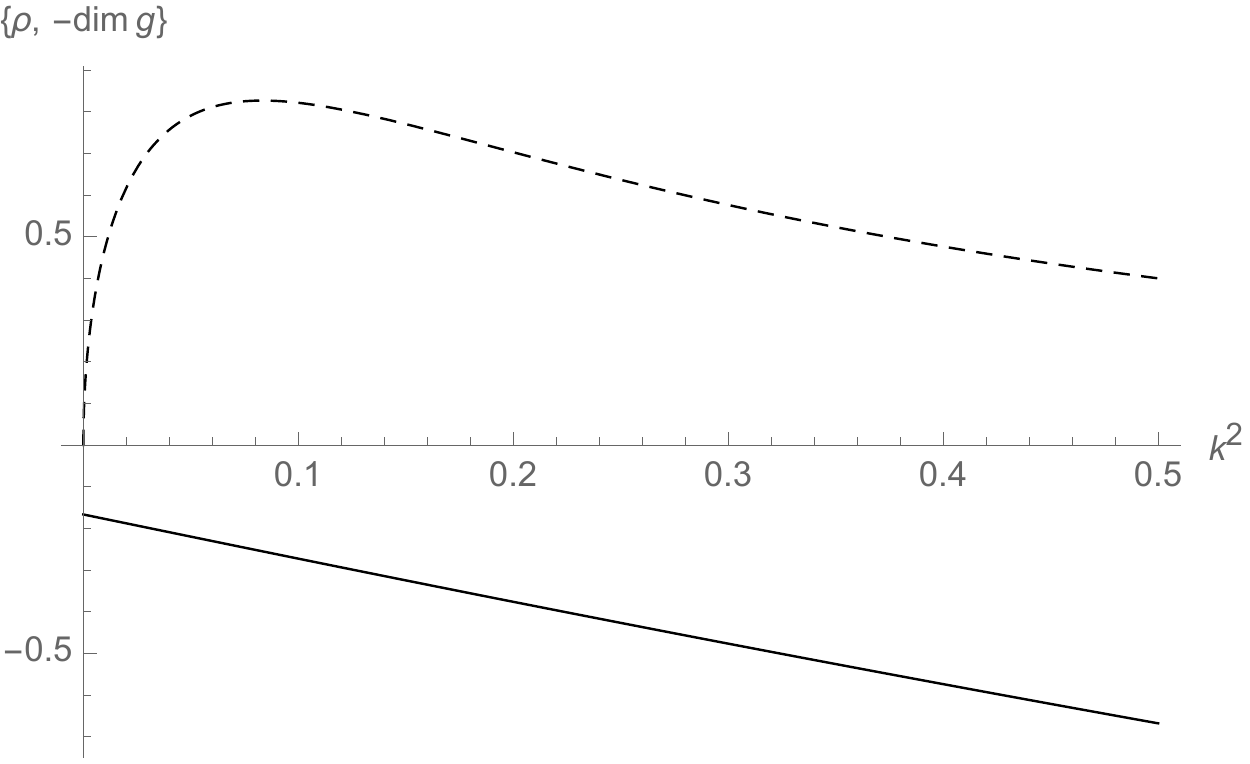}
\includegraphics[scale=0.55]{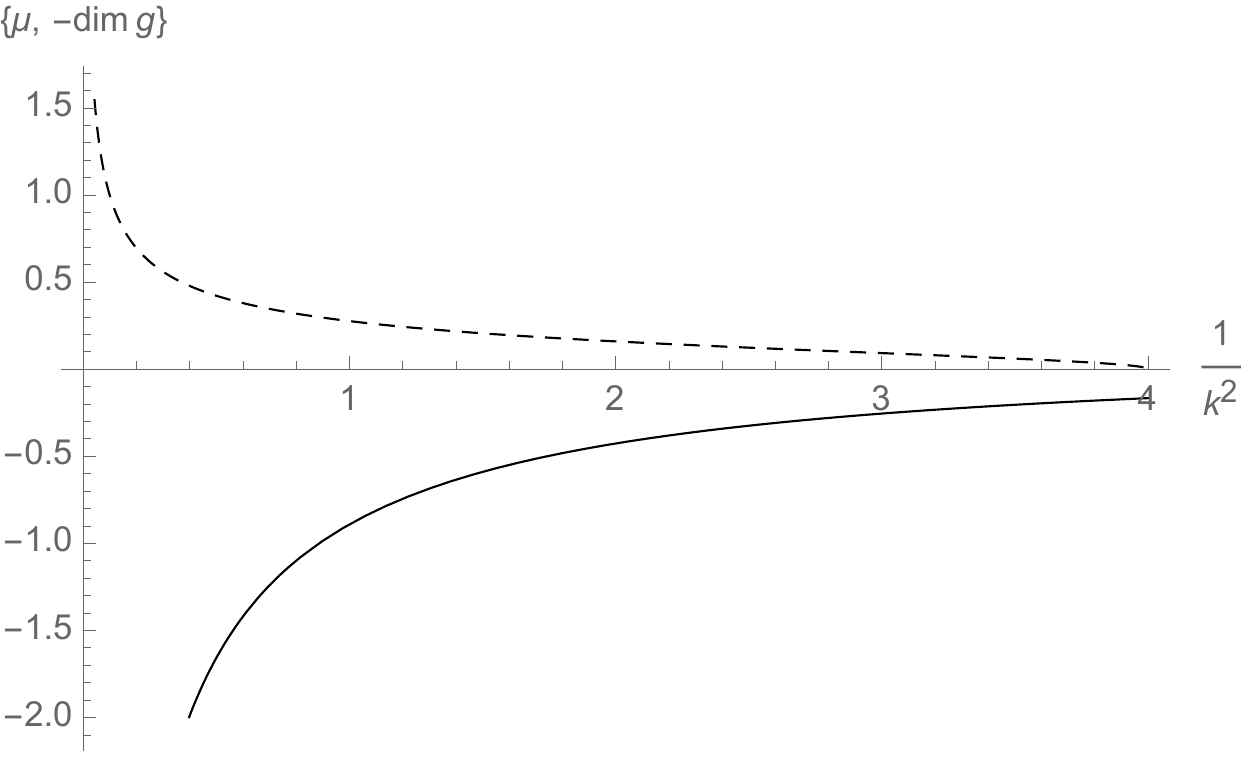}
\caption{Upstairs: Canonical dimension (solid line) versus the inverse MP distribution (the dashed line). Downstairs: The canonical dimension (solid line) versus the MP distrubution (dashed line).} \label{MP2}
\end{figure}

As a first observation, the flow equations involve loop integrals of type:
\begin{equation}
L:=\int_0^k \tilde\rho(p^2)p dp\,,
\end{equation}
which, for large $k$, and $\tilde{\rho}(p^2)\propto (p^2)^\alpha$ behaves like $k^{2\alpha+2}$, as computed in the previous section. Therefore:
\begin{equation}
\ln L = (2\alpha+2)\ln(k)+C\,,
\end{equation}
For $C$ being a numerical constant depending on $\alpha$, and $\ln(k)=t$, the scale parameter along the RG flow called abusively \textit{time}. Therefore, for a power law distribution, $\tau:= \ln L$ and $t$ are related by an affine transformation; $d\tau$ and $dt$ being proportional. The two times are, with this respect, essentially physically equivalents. Obviously, such a relation break down for arbitrary distributions. However, flow equations simplify using $\tau$ rather than $t$. Let us consider the flow of $u_2$:
\begin{equation}
\dot{u}_2=- \frac{2gk^2}{(k^2+u_2)^2} \int_0^k \tilde{\rho}(p^2)p dp\,.
\end{equation}
Computing the derivative:
\begin{equation}
\frac{d\tau}{dt}=k^2 \tilde\rho(k^2) \frac{1}{ \int_0^k \tilde{\rho} (p^2)p dp} \,,
\end{equation}
we get:
\begin{equation}
\dot{u}_2=- \frac{2g}{(1+\bar{u}_2)^2} \tilde{\rho}(k^2) \frac{dt}{d\tau} \,,
\end{equation}
where we used the fact that $dt/d\tau=(d\tau/dt)^{-1}$. Multiplying term by term with $dt/d\tau$, we then get:
\begin{equation}
\frac{d \bar{u}_2}{d\tau}=-2 \frac{dt}{d\tau}\bar{u}_2-\frac{2g}{(1+\bar{u}_2)^2} \frac{\tilde{\rho}(k^2)}{k^2} \left(\frac{dt}{d\tau}\right)^2\,.
\end{equation}
Therefore, defining the dimensionless coupling $\bar{g}$ as:
\begin{equation}
g \frac{\tilde{\rho}(k^2)}{k^2} \left(\frac{dt}{d\tau}\right)^2 =:\bar{g}\,,\label{barg}
\end{equation}
we obtain finally:
\begin{equation}
\frac{d \bar{u}_2}{d\tau}=-2 \frac{dt}{d\tau}\bar{u}_2- \frac{2\bar{g}}{(1+\bar{u}_2)^2} \,.
\end{equation}
In the same way, for the coupling, we get:
\begin{equation}
\frac{dg}{d\tau}= \frac{12g^2}{(1+\bar{u}_2)^3}\,\frac{\tilde{\rho}(k^2)}{k^2} \left(\frac{dt}{d\tau}\right)^2\,.
\end{equation}
We have to write the equation for the dimensionless coupling $\bar{g}$, therefore, we need to compute the derivative of the dimensionless coupling given by \eqref{barg}:
\begin{equation}
{\bar{g}}^\prime={g}^\prime \frac{\rho(k^2)}{k^2} \left(\frac{dt}{d\tau}\right)^2+2\bar{g}\left(\frac{t^{\prime\prime}}{t^\prime}+t^\prime\left(\frac{1}{2} \frac{d \ln\tilde{\rho}}{dt}-1\right)\right)\,.
\end{equation}
Therefore:
\begin{equation}
\frac{d\bar{g}}{d\tau}=2\bar{g}\left(\frac{t^{\prime\prime}}{t^\prime}+t^\prime\left(\frac{1}{2} \frac{d \ln\tilde{\rho}}{dt}-1\right)\right)+ \frac{12\bar{g}^2}{(1+\bar{u}_2)^3}\,,
\end{equation}
where:
\begin{equation}
X^\prime := \frac{d X}{d \tau}\,.
\end{equation}
The role played bu the canonical dimension is now played by a more complicated function, defining scale by scale:
\begin{equation}
-\dim(g):=2\left(\frac{t^{\prime\prime}}{t^\prime}+t^\prime\left(\frac{1}{2} \frac{d \ln\tilde{\rho}}{dt}-1\right)\right)\,,
\end{equation}
which is pictured in Figure \ref{MP2} for the MP distribution. The dimension is positive everywhere, meaning, as expected that the Gaussian fixed point will be stable placing the cut-off at an arbitrary scale. This scale invariance may be viewed as another characterization of noise, and the MP law provides the common analytic representation of noisy signals. The question is now, what happens when the MP law is disturbed by a macroscopic signal? One expects that, for a sufficiently big signal, the canonical dimension will become negative from a certain scale, breaking the scale invariance. As we will see in the next section, this is precisely what happens.

\section{Numerical investigation}\label{numinv}

\begin{figure}
\includegraphics[scale=0.24]{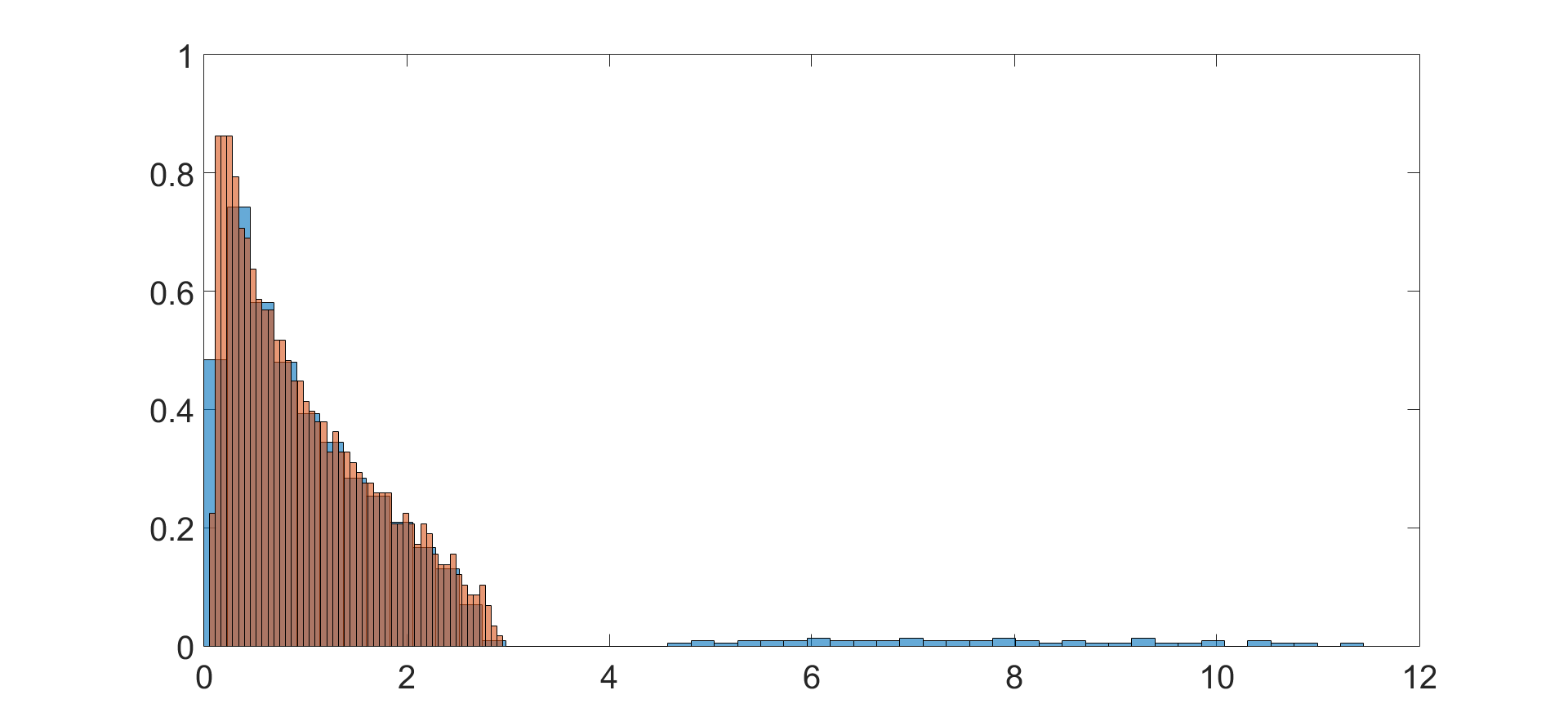} \caption{Red Histogram: data set with i.i.d random entries for $p = 1000, N=2000$. Blue histogram: perturbation of the random distribution with a matrix of rank $k=50$ (defining the size of the signal) and ratio $x = 300$.} \label{fig6}
\end{figure}

In this section, as announced, we investigate data sets from an RG point of view.  We focus on artificial sets, made of some constant spikes disturbed by a random signal materializing with a matrix with random entries, playing the role of the noise. Figure \ref{fig6} provide a typical spectrum, for purely random entries (the red histogram) and when a non-random signal is added (the blue histogram). In this simple case, the standard PCA could be applied, the largest eigenvalues being far from the bulk, which tends toward the MP distribution for large $p$ and $N$. The distinction between signal and noise is therefore clear for this example. This is however not the case for the spectrum given in Figure \ref{fig7} below.
In this Figure, there is no clean separation between what is information and what is noise, the interesting part of the signal being merged onto the noise. Figure \ref{fig8} provides the canonical dimension with and without signal. For the case without signal, we recover the same curve as for the MP law, see Figure \ref{MP2}, up to some irrelevant irregularities due to the numerical interpolation. Indeed, the computation of the canonical dimension requires first and second derivative, which are very sensitive to the sharp variations of the interpolation. Adding the signal, we show that the canonical dimension is changed, and increase significantly in the region of large eigenvalues, to become positive from a certain scale. The interpretation of the phenomena is clear from the analysis of section \ref{powerlaw}. For very small eigenvalues, the canonical dimension is essentially unaffected by the signal, and the curve is the same as the one without the signal. Moving toward the large eigenvalue region, however, the deformation increase and the canonical dimension becomes larger, meaning, in the point of view of the RG that system reaches the critical region. Finally, the canonical dimension becomes positive, and the Gaussian fixed point becomes unstable. This illustrate what we expected in the discussion of Section \ref{sec2}. The universal properties of the large scale distributions are affected by the presence of the signal up to a certain range; which we can ‘‘detect" by the behavior of the RG flow. Note that this observation seems to contradict the assumptions justifying the perturbative treatment in \cite{pca0}. Indeed, what we show is that a purely noisy signal is not suitably described by a Gaussian distribution. This can be the case however if the strength of the signal makes the scaling dimension for the coupling negative.
\medskip

Interestingly, the canonical dimension becomes positive around $2.2$, well before the theoretical end of the MP distribution. At this stage, the signal prevails over the noise. However, the competition between them starts before this point. From the point where the signal goes out, the estimated size for the signal is $52$, which have to be compared to the number $65$ of eigenvalues for the input signal. Then, in this example, the method allows recovering $80$ per cent of the original signal.
\begin{figure}
\includegraphics[scale=0.23]{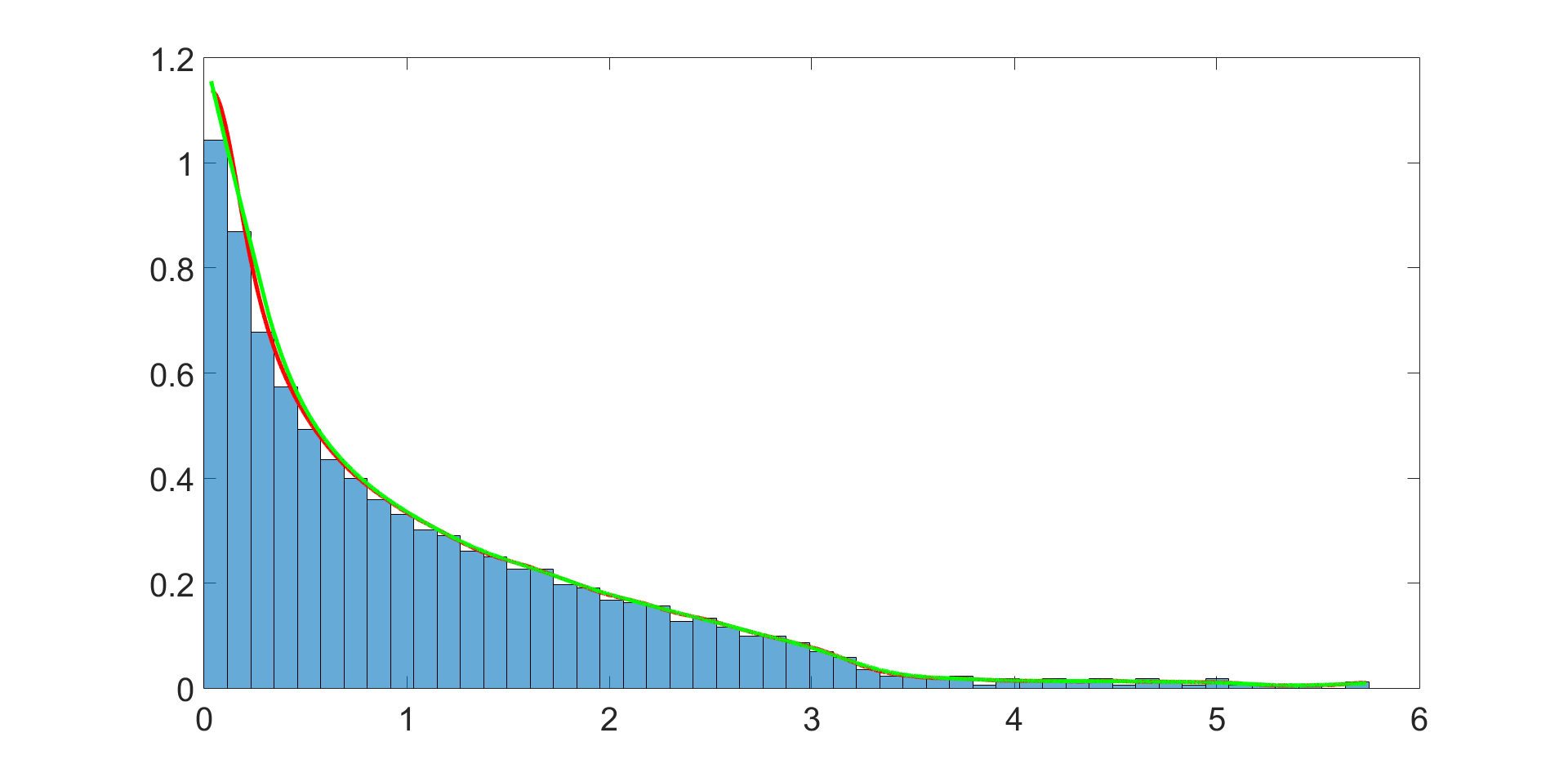}
\caption{Eigenvalue distribution for $p=1500$, $N=2000$; and the eigenvalues of the matrix playing the role of the signal distributed with different weights. For this case, there is no clean separation between noisy and relevant degrees of freedom. The green curve corresponds to a numerical interpolation of the discrete distribution.}\label{fig7}
\end{figure}

\begin{figure}
\includegraphics[scale=0.22]{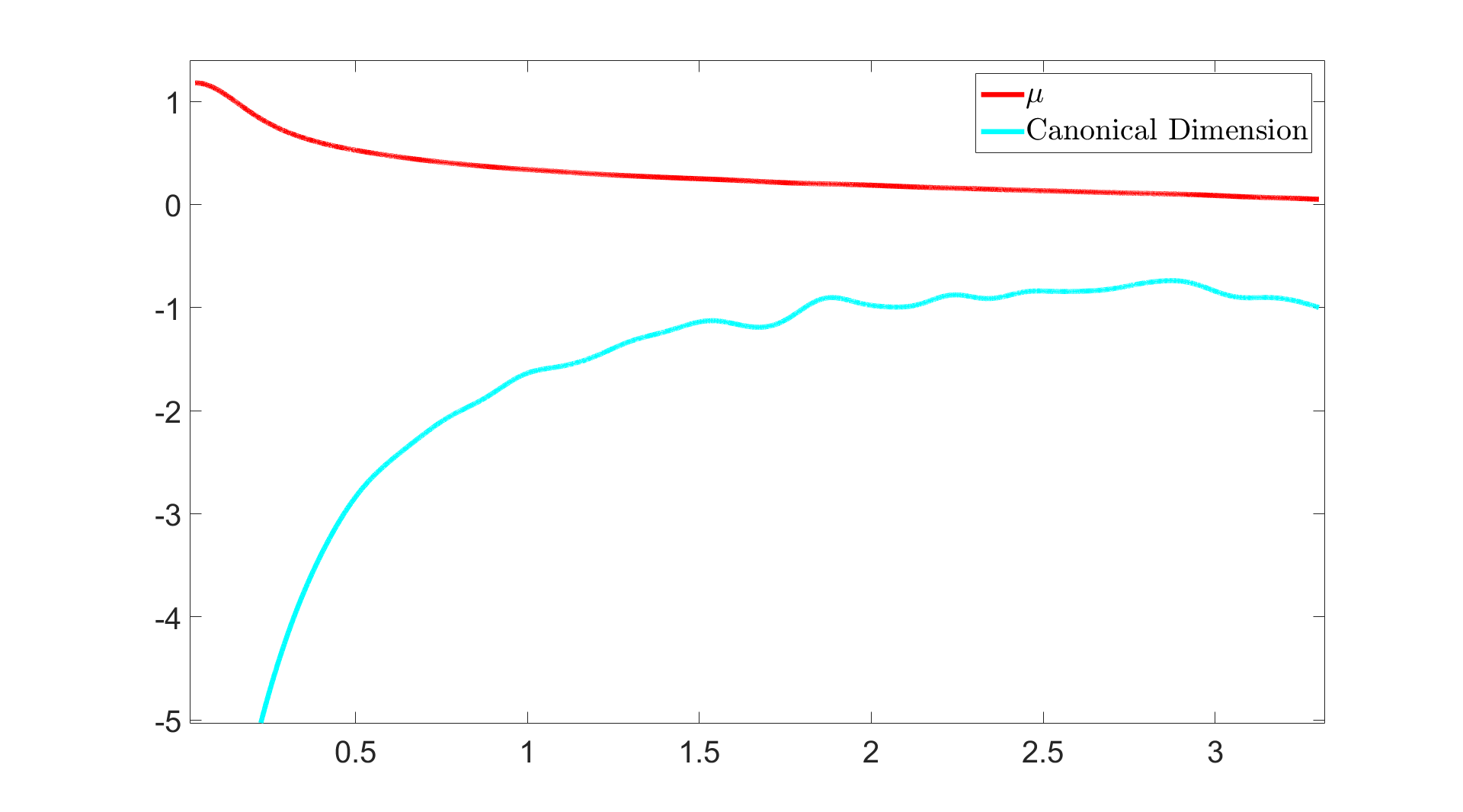} \includegraphics[scale=0.22]{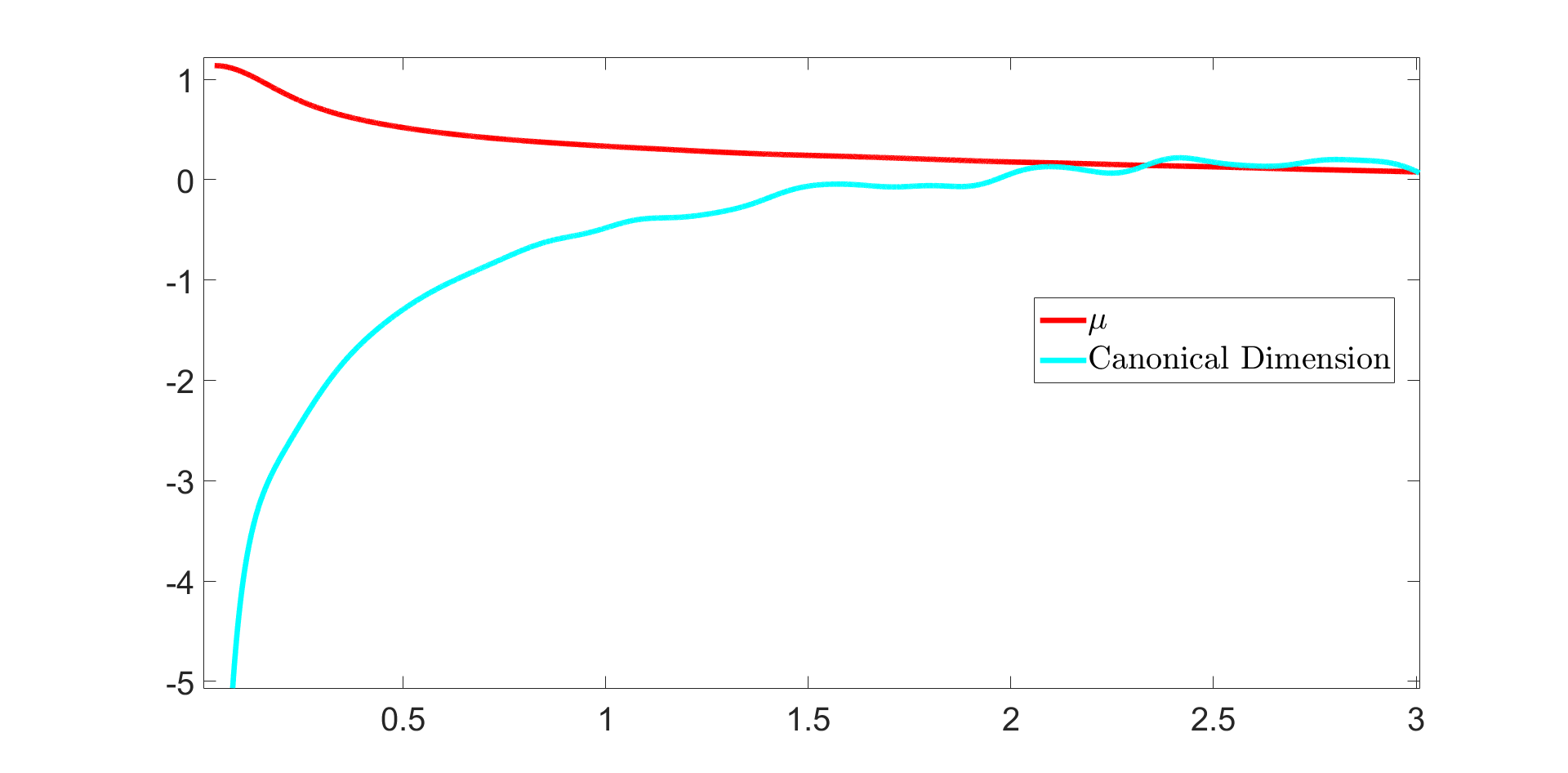}
\caption{The canonical dimension version the eigenvalue distribution. Without signal (upstairs) and with signal (downstairs).}\label{fig8}
\end{figure}

\section{Conclusion and perspectives}\label{sec5}

In this paper, we continue the idea developed in \cite{pca0} of a connection between RG and PCA. The fact that both PCA and RG search for simplifying systems involving a large degree of freedom was an indication that RG technique could be applied in the PCA context when standard methods break down. This is especially the case for continuous spectra, where RG is expected able to clarify the separation between data and noise. The separation, in the RG language, plays the role of a cut-off over degrees of freedom and varying it, we expect that noisy and relevant degrees of freedom distinguish from their influence on the RG behavior. \\
In this paper, we introduced a nonperturbative framework, and discuss arbitrary spectra from a criterion involving a generalization of the so-called canonical dimension. This canonical dimension is positive for MP distribution, but as we have seen considering artificial data sets, it may be strongly influenced and can become positive when a signal is added to the noise. These results added to the ones of \cite{pca0}-\cite{bradde2019} show that RG may be fruitfully used as a promising tool in the PCA context for challenging problems. \\
Surely, some questions remain open and have to be clarified in the future. First of all, the criteria providing by the dimension requires to compute of the first and second derivative, which are not easy to estimate from discrete spectra. In the analysis of section \ref{numinv}, we used numerical tools to improve the analyticity of the interpolation curve, discarding the most singular points. A finer numerical technique has to be used to improve this point. Secondly, our approximations are limited to the simpler truncation, and deeper investigations of the theory space could provide finer arguments to clarify the cut-off between noisy and relevant degrees of freedom. For instance, one can imagine that higher truncations could reveal finer details over data, breaking scale invariance at higher scales. Another aspect concern the role of phase transitions, where a very large number of microscopic degrees of freedom behave collectively to generate macroscopic effects. This is a common feature in physical systems involving many degrees of freedom, and our analysis seems to indicate that approaching the separation point between information and noise the behavior of the RG flow is very reminiscent of a system near criticality. One may expect that such an effective description allows improving the criteria, reasoning over the efficient dimensions near the non-Gaussian fixed point rather than the canonical dimensions, only valid in the Gaussian region.

\onecolumngrid


\begin{thebibliography}{99}

\bibitem{Avdoshkin:2019trj} 
  A.~Avdoshkin and A.~Dymarsky,
  ``Euclidean operator growth and quantum chaos,''
  arXiv:1911.09672 [cond-mat.stat-mech].

\bibitem{Sunderhauf:2019djv} 
  C.~Sünderhauf, L.~Piroli, X.~L.~Qi, N.~Schuch and J.~I.~Cirac,
  ``Quantum chaos in the Brownian SYK model with large finite $N$: OTOCs and tripartite information,''
  JHEP {\bf 1911}, 038 (2019)
  doi:10.1007/JHEP11(2019)038
  [arXiv:1908.00775 [quant-ph]].


\bibitem{Franz:1998nt} 
  S.~Franz, M.~Mézard, G.~Parisi and L.~Peliti,
  ``Measuring equilibrium properties in aging systems,''
  Phys.\ Rev.\ Lett.\  {\bf 81}, 1758 (1998)
  doi:10.1103/PhysRevLett.81.1758
  [cond-mat/9803108 [cond-mat.stat-mech]].


\bibitem{Bouchaud:1997uv} 
  J.~P.~Bouchaud, L.~F.~Cugliandolo, J.~Kurchan and M.~Mezard,
  ``Out of equilibrium dynamics in spin-glasses and other glassy systems,''
  cond-mat/9702070 [cond-mat.dis-nn].

\bibitem{Bouchaud:1995wx} 
  J.~P.~Bouchaud, L.~Cugliandolo, J.~Kurchan and M.~Mezard,
  ``Mode coupling approximations, glass theory and disordered systems,''
  Physica A {\bf 226}, 243 (1996)
  doi:10.1016/0378-4371(95)00423-8
  [cond-mat/9511042].

\bibitem{Mezard:1988dv} 
  M.~Mezard and G.~Parisi,
  ``The Euclidean Matching Problem,''
  LPTENS-88/21.
  
  
\bibitem{Krauth:1987vr} 
  W.~Krauth and M.~Mezard,
  ``Learning Algorithms With Optimal Stability In Neural Networks,''
  J.\ Phys.\ A {\bf 20}, L745 (1987).
  doi:10.1088/0305-4470/20/11/013
  
  
\bibitem{Carleo:2019ptp} 
  G.~Carleo, I.~Cirac, K.~Cranmer, L.~Daudet, M.~Schuld, N.~Tishby, L.~Vogt-Maranto and L.~Zdeborová,
  ``Machine learning and the physical sciences,''
  arXiv:1903.10563 [physics.comp-ph].


\bibitem{Loebl:2005pk} 
  M.~Loebl and L.~Zdeborova,
  ``The 3-D dimer and Ising problems revisited,''
  Eur.\ J.\ Combinatorics {\bf 29}, 966 (2008)
  doi:10.1016/j.ejc.2007.11.013
  [cond-mat/0505384].

\bibitem{Charbonneau:2016uvz} 
  P.~Charbonneau and S.~Yaida,
  ``Nontrivial critical fixed point for replica-symmetry-breaking transitions,''
  Phys.\ Rev.\ Lett.\  {\bf 118}, no. 21, 215701 (2017)
  doi:10.1103/PhysRevLett.118.215701
  [arXiv:1607.04217 [cond-mat.stat-mech]].

\bibitem{Yeo:2012ur} 
  J.~H.~Yeo and M.~A.~Moore,
  ``Renormalization group analysis of the M-p-spin glass model with p=3 and M = 3,''
  Phys.\ Rev.\ B {\bf 85}, 100405 (2012)
  doi:10.1103/PhysRevB.85.100405
  [1111.3105 [cond-mat.stat-mech]].

\bibitem{Castellana:2010zz} 
  M.~Castellana and G.~Parisi,
  ``Renormalization group computation of the critical exponents of hierarchical spin glasses,''
  Phys.\ Rev.\ E {\bf 82}, 040105 (2010)
  doi:10.1103/PhysRevE.82.040105
  [arXiv:1006.5628 [cond-mat.dis-nn]].

\bibitem{Pezzella:1997nu} 
  U.~Pezzella and A.~Coniglio,
  ``Spin glasses and frustrated percolation: a renormalization group approach,''
  Physica A {\bf 237}, 353 (1997).

\bibitem{Dotsenko:1987zc} 
  V.~S.~Dotsenko,
  ``Towards A Renormalization Group Theory Of Spin Glasses,''
  IC/87/154.


\bibitem{Collet:1983dmw} 
  P.~Collet, J.~P.~Eckmann, V.~J.~Glaser and A.~Martin,
  ``A spin glass with random couplings,''
  J.\ Statist.\ Phys.\  {\bf 36}, 89 (1984).
  doi:10.1007/BF01015728

\bibitem{solo2}
E.~Aygun, A.~Erzan, ``Spectral renormalization group theory on networks'' J. Phys Conf Series  {\bf 319}, 012007
(2011).




\bibitem{dlearning1}
P.~Mehta, D.~Schwab, ``An exact mapping between the variational renormalization group and deep learning''
arXiv:1410.3831 [stat.ML].



\bibitem{Hattori:1987jm} 
  K.~Hattori, T.~Hattori and H.~Watanabe,
  ``Gaussian Field Theories On General Networks And The Spectral Dimensions,''
  Prog.\ Theor.\ Phys.\ Suppl.\  {\bf 92}, 108 (1987).
  doi:10.1143/PTPS.92.108.


\bibitem{pca0}
S.~Bradde, W.~Bialek,``PCA meets RG''  Journal of Statistical Physics, {\bf 167}, Issue 3–4, pp 462–475, (2017) doi.org/10.1007/s10955-017-1770-6	arXiv:1610.09733 [physics.bio-ph].

\bibitem{pca1}
J.~Shlens, ``A tutorial on principal components analysis'', arXiv:1404.1100 [cs.LG] (2014).

\bibitem{bradde2019}
S.~Bradde,  A.~Nourmohammad,  S.~Goyal, V.~Balasubramanian, ``The size of the immune repertoire of bacteria'' arXiv:1903.00504 [q-bio.PE] (2019).


\bibitem{Bao:2019hfc} 
  C.~Bao,
  ``Loop Optimization of Tensor Network Renormalization: Algorithms and Applications,''

 
 
\bibitem{Campeti:2019ylm} 
  P.~Campeti, D.~Poletti and C.~Baccigalupi,
  ``Principal component analysis of the primordial tensor power spectrum,''
  JCAP {\bf 1909}, no. 09, 055 (2019)
  doi:10.1088/1475-7516/2019/09/055
  [arXiv:1905.08200 [astro-ph.CO]].


\bibitem{Woloshyn:2019oww} 
  R.~M.~Woloshyn,
  ``Learning phase transitions: comparing PCA and SVM,''
  arXiv:1905.08220 [cond-mat.stat-mech].
  
  
 \bibitem{PCA1}
JP.~Benzécri,  ``Analyse des données. T2 (leçons sur l'analyse factorielle et la reconnaissance des formes et travaux du Laboratoire de statistique de l'Université de Paris 6. T. 2 : l'analyse des correspondances),'' Dunod Paris Bruxelles Montréal, 1973.
 
 \bibitem{PCA2}
 H.~Hotelling,  ``Analysis of a complex of statistical variables into principal components,''   Journal of Educational Psychology, {\bf 24}, 417–441, and 498–520   (1933).
 
 \bibitem{PCA3}
 H.~Abdi, LJ.~Williams, ``Principal component analysis". Wiley Interdisciplinary Reviews,''   Computational Statistics. {\bf 2} (4): 433–459. arXiv:1108.4372. doi:10.1002/wics.101    (2010).
 
 \bibitem{PCA4}
Haiping.~Lu,   KN.~Plataniotis,  AN.~Venetsanopoulos, ``A Survey of Multilinear Subspace Learning for Tensor Data,''    Pattern Recognition. {\bf 44} (7): 1540–1551. doi:10.1016/j.patcog.2011.01.004    (2011).

\bibitem{PCA5}
G.~Yue, Dy.~Jennifer,  ``Sparse Probabilistic Principal Component Analysi,'' Journal of Machine Learning Research Workshop and Conference Proceedings   (2009).
  
\bibitem{Foreman:2018qei} 
  S.~Foreman, J.~Giedt, Y.~Meurice and J.~Unmuth-Yockey,
  ``Machine learning inspired analysis of the Ising model transition,''
  PoS LATTICE {\bf 2018}, 245 (2018).
  doi:10.22323/1.334.0245


\bibitem{Beny:2018agy} 
  C.~Bény,
  ``Inferring relevant features: From QFT to PCA,''
  Int.\ J.\ Quant.\ Inf.\  {\bf 16}, no. 08, 1840012 (2018)
  doi:10.1142/S0219749918400129
  [arXiv:1802.05756 [cs.LG]].



\bibitem{Foreman:2017mbc} 
  S.~Foreman, J.~Giedt, Y.~Meurice and J.~Unmuth-Yockey,
  ``RG inspired Machine Learning for lattice field theory,''
  EPJ Web Conf.\  {\bf 175}, 11025 (2018)
  doi:10.1051/epjconf/201817511025
  [arXiv:1710.02079 [hep-lat]].


\bibitem{KADANOFF:1967zz} 
  L.~P.~Kadanoff {\it et al.},
  ``Static Phenomena Near Critical Points: Theory and Experiment,''
  Rev.\ Mod.\ Phys.\  {\bf 39}, 395 (1967).
  doi:10.1103/RevModPhys.39.395
  
  
\bibitem{Sokal:1994un} 
  A.~D.~Sokal, A.~C.~D.~van Enter and R.~Fernandez,
  ``Regularity properties and pathologies of position space renormalization group transformations: Scope and limitations of Gibbsian theory,''
  J.\ Statist.\ Phys.\  {\bf 72}, 879 (1994)
  doi:10.1007/BF01048183
  [hep-lat/9210032].

\bibitem{Kadanoff:2009ns} 
  L.~P.~Kadanoff,
  ``More is the Same: Phase Transitions and Mean Field Theories,''
  J.\ Statist.\ Phys.\  {\bf 137}, 777 (2009)
  doi:10.1007/s10955-009-9814-1
  [arXiv:0906.0653 [physics.hist-ph]].

\bibitem{Wilson:1971dc} 
  K.~G.~Wilson and M.~E.~Fisher,
  ``Critical exponents in 3.99 dimensions,''
  Phys.\ Rev.\ Lett.\  {\bf 28}, 240 (1972).
  doi:10.1103/PhysRevLett.28.240
  
  
\bibitem{Wetterich:1991be} 
  C.~Wetterich,
``The Average action for scalar fields near phase transitions,''
  Z.\ Phys.\ C {\bf 57}, 451 (1993).
  doi:10.1007/BF01474340



\bibitem{Wetterich:1992yh} 
  C.~Wetterich,
  ``Exact evolution equation for the effective potential,''
  Phys.\ Lett.\ B {\bf 301}, 90 (1993)
  doi:10.1016/0370-2693(93)90726-X
  [arXiv:1710.05815 [hep-th]].

\bibitem{Litim:2000ci} 
  D.~F.~Litim,
  ``Optimization of the exact renormalization group,''
  Phys.\ Lett.\ B {\bf 486}, 92 (2000)
 doi:10.1016/S0370-2693(00)00748-6
  [hep-th/0005245].
  


\bibitem{Litim:2001dt} 
  D.~F.~Litim,
  ``Derivative expansion and renormalization group flows,''
  JHEP {\bf 0111}, 059 (2001)
  doi:10.1088/1126-6708/2001/11/059
  [hep-th/0111159].
  
  
\bibitem{Canet:2002gs} 
  L.~Canet, B.~Delamotte, D.~Mouhanna and J.~Vidal,
  ``Optimization of the derivative expansion in the nonperturbative renormalization group,''
  Phys.\ Rev.\ D {\bf 67}, 065004 (2003)
  doi:10.1103/PhysRevD.67.065004
  [hep-th/0211055].
  
\bibitem{Delamotte:2007pf} 
  B.~Delamotte,
  ``An Introduction to the nonperturbative renormalization group,''
  Lect.\ Notes Phys.\  {\bf 852}, 49 (2012)
  doi:10.1007/978-3-642-27320-$9_2$
  [cond-mat/0702365 [cond-mat.stat-mech]].


\bibitem{Berges:2000ew} 
  J.~Berges, N.~Tetradis and C.~Wetterich,
  ``Nonperturbative renormalization flow in quantum field theory and statistical physics,''
  Phys.\ Rept.\  {\bf 363}, 223 (2002)
  doi:10.1016/S0370-1573(01)00098-9
  [hep-ph/0005122].
  
\bibitem{Gies:2001nw} 
  H.~Gies and C.~Wetterich,
  ``Renormalization flow of bound states,''
  Phys.\ Rev.\ D {\bf 65}, 065001 (2002)
  doi:10.1103/PhysRevD.65.065001
  [hep-th/0107221].
  
 

\bibitem{Reuter:2011ah} 
  M.~Reuter and F.~Saueressig,
  JHEP {\bf 1112}, 012 (2011)
  doi:10.1007/JHEP12(2011)012
  [arXiv:1110.5224 [hep-th]].


\bibitem{Pawlowski:2015mlf} 
  J.~M.~Pawlowski, M.~M.~Scherer, R.~Schmidt and S.~J.~Wetzel,
  ``Physics and the choice of regulators in functional renormalisation group flows,''
  Annals Phys.\  {\bf 384}, 165 (2017)
  doi:10.1016/j.aop.2017.06.017
  [arXiv:1512.03598 [hep-th]].


\bibitem{finan1}
M.~Marsili, ``Dissecting financial markets: sectors and states,'' Quantitative Finance {\bf 2}, 297-302 (2002).


\bibitem{finan2}
F.~Lillo and RN.~Mantegna, ``Variety and violatility in financial markets,''. Phys Rev E {\bf 62}, 6126-6134  (2000).
  
 \bibitem{finan3}
 KS.~Brown, CC.~Hill, GA.~Calero, CR.~Myers, KH.~Lee, JP.~Sethna, and RA.~Cerione, ``The statistical mechanics ofcomplex signaling networks: Nerve growth factor aignaling,''  Phys Biol  {\bf 1}, 184–195 (2004).

  \bibitem{finan4}
  RN.~Gutenkunst, JJ.~Waterfall, FP.~Casey, KS.~Brown, CR.~Myers,  JP.~Sethna, ``Universally sloppy parameter sensitivities in systems biology,''  PLoS Comput Biol {\bf 3}, e189 (2007).
  
  \bibitem{finan5}
  MK.~Transtrum, BB.~Machta,  JP.~Sethna, ``Geometry of nonlinear least squares with applications to sloppy models and optimization,'' Phys Rev E {\bf 83}, 036701 (2011).
  
  \bibitem{finan6}
  BB.~Matcha, R.~Chachra, MK.~Transtrum, JP.~Sethna,  ``Parameter space compression underlies emergent theoriesand predictive models,'' Science {\bf 342}, 604–607 (2013).
  
  \bibitem{finan7}
  JJ.~Waterfall, FP.~Casey, RN.~Gutenkunst, KS.~Brown, CR.~Myers, PW.~Brouwer, V.~Elser, and JP.~Sethna, ``Sloppymodel universality class and the Vandermonde matrix,''  Phys Rev Lett {\bf 97},  150601 (2006).
  
  \bibitem{finan8}
  JP.~Bouchaud, M.~Potters, ``Financial applications. InThe Oxford Handbook of Random Matrix Theory.''  G.~Ake-mann, J.~Baik,  P.~Di Francesco, eds (Oxford University Press, 2011); arXiv:0910.1205 [q–fin.ST] (2009).
 
 
\bibitem{Morampudi:2018iev} 
  S.~C.~Morampudi, A.~Chandran and C.~R.~Laumann,
  ``Universal entanglement of typical states in constrained systems,''
  arXiv:1810.04157 [quant-ph].
  
  
\bibitem{Kanzieper:2010xr} 
  E.~Kanzieper and N.~Singh,
  ``Non-Hermitean Wishart random matrices (I),''
  J.\ Math.\ Phys.\  {\bf 51}, 103510 (2010)
  doi:10.1063/1.3483455
  [arXiv:1006.3096 [math-ph]].
  
  

\bibitem{Lu:2014jua} 
  X.~Lu and H.~Murayama,
  ``Universal Asymptotic Eigenvalue Distribution of Large $N$ Random Matrices --- A Direct Diagrammatic Proof to Marchenko-Pastur Law ---,''
  arXiv:1410.3503 [hep-th].

  
\end{thebibliography}
\end{document}